\documentclass{article}
\usepackage{}
\usepackage{epsfig}

\usepackage{a4}

\date{\today} 

\usepackage{epsfig}
\usepackage{amsmath}
\usepackage{amsfonts} 
\usepackage{amssymb}
\usepackage{ifthen}

 \usepackage{etex}
 \reserveinserts{18}

\begin{document}
\title{$SO(2)$ gauged Skyrmions in $4+1$ dimensions }
\author{{\large Francisco Navarro-L\'erida,}$^{1}$
{\large Eugen Radu}$^{2}$
and {\large D. H. Tchrakian}$^{3,4}$ 
\\ 
\\
$^{1}${\small  Departamento de F\'isica Te\'orica and IPARCOS, Ciencias F\'isicas,}\\
{\small Universidad Complutense de Madrid, E-28040 Madrid, Spain}\\   
$^{2}${\small
 Departamento de Matem\'atica da Universidade de Aveiro and }
\\
{\small Center for Research and Development in Mathematics and Applications,}
\\
{\small 
 Campus de Santiago, 3810-183 Aveiro, Portugal
}\\ 
$^{3}${\small School of Theoretical Physics, Dublin Institute for Advanced Studies,}
\\{\small 10 Burlington Road, Dublin 4, Ireland }
\\
$^{4}${\small  Department of Computer Science, NUI Maynooth, Maynooth, Ireland}}

\date{\today}
\newcommand{\dd}{\mbox{d}}
\newcommand{\tr}{\mbox{tr}}
\newcommand{\la}{\lambda}
\newcommand{\om}{\omega}
\newcommand{\ka}{\kappa}
\newcommand{\ta}{\theta}
\newcommand{\f}{\phi}
\newcommand{\vf}{\varphi}
\newcommand{\vr}{\varrho}
\newcommand{\F}{\Phi}
\newcommand{\al}{\alpha}
\newcommand{\bt}{\beta}
\newcommand{\ga}{\gamma}
\newcommand{\de}{\delta}
\newcommand{\si}{\sigma}
\newcommand{\Si}{\Sigma}
\newcommand{\bomega}{\mbox{\boldmath $\omega$}}
\newcommand{\bOmega}{\mbox{\boldmath $\Omega$}}
\newcommand{\bsi}{\mbox{\boldmath $\sigma$}}
\newcommand{\bchi}{\mbox{\boldmath $\chi$}}
\newcommand{\bal}{\mbox{\boldmath $\alpha$}}
\newcommand{\bpsi}{\mbox{\boldmath $\psi$}}
\newcommand{\brho}{\mbox{\boldmath $\varrho$}}
\newcommand{\beps}{\mbox{\boldmath $\varepsilon$}}
\newcommand{\bxi}{\mbox{\boldmath $\xi$}}
\newcommand{\bbeta}{\mbox{\boldmath $\beta$}}
\newcommand{\ee}{\end{equation}}
\newcommand{\eea}{\end{eqnarray}}
\newcommand{\be}{\begin{equation}}
\newcommand{\bea}{\begin{eqnarray}}

\newcommand{\ii}{\mbox{i}}
\newcommand{\e}{\mbox{e}}
\newcommand{\pa}{\partial}
\newcommand{\Om}{\Omega}
\newcommand{\vep}{\varepsilon}
\newcommand{\bfph}{{\bf \phi}}
\def\theequation{\arabic{equation}}
\renewcommand{\thefootnote}{\fnsymbol{footnote}}
\newcommand{\re}[1]{(\ref{#1})}

\newcommand{\eins}{1\hspace{-0.56ex}{\rm I}}
\newcommand{\R}{\mathbb R}
\newcommand{\C}{\mathbb C}
\newcommand{\p}{\mathbb P}
\renewcommand{\thefootnote}{\arabic{footnote}}

\maketitle


\bigskip

\begin{abstract}
 We study the simplest $SO(2)$ gauged $O(5)$ Skyrme models in $4+1$ (flat) dimensions. 
In the gauge decoupled limit, the model supports topologically stable solitons (Skyrmions)
and after gauging, the static energy of the solutions is bounded from below by a ``baryon number''. 
The studied model features both Maxwell and Maxwell--Chern-Simons dynamics. 
The considered configurations are subject to bi-azimuthal symmetry in the $\R^4$ 
subspace resulting in a two dimensional subsystem, as well as subject to an enhanced symmetry relating the two
planes in the $\R^4$ subspace, which results in a one dimensional subsystem. 
Numerical solutions are constructed in both cases. 
In the purely magnetic case, fully bi-azimuthal solutions were given,
while electrically charged and spinning solutions 
were constructed only in the radial (enhanced symmetric) case, 
both in the presence of a Chern-Simons term, and in its absence.
We find that,  in contrast with the analogous models in $2+1$ dimensions,
the presence of the Chern-Simons term in the model  under study here  results only in quantitative effects. 
\end{abstract}
\medskip
\medskip

\tableofcontents

\section{Introduction}
The gauging of the Skyrmion, namely of the soliton of the $O(4)$ sigma model on $\R^3$, is recognised to be of physical relevance in the study of the electrically charged nucleon. This was
considered by Callan and Witten~\cite{Callan:1983nx} in the context of baryon number violation.
Gauging a Skyrme (sigma model) scalar results in the deformation of the lower bound on the energy, which prior to gauging is the topological charge, namely the winding number.
The most prominent such example is the $U(1)$, or $SO(2)$, gauged Skyrme system in $3+1$ dimensions, 
the earliest work being \cite{Callan:1983nx}, where the emphasis was on baryon number violation, and subsequently \cite{Piette:1997ny}, where the dependence of the mass
of the proton on the electric charge was studied, and \cite{Radu:2005jp}, 
where the spin of the proton was considered. 
The gauging prescription used in \cite{Piette:1997ny} and \cite{Radu:2005jp} 
coincides with that used in \cite{Callan:1983nx}.

These studies, \cite{Callan:1983nx,Piette:1997ny,Radu:2005jp}, apply to the $SO(2)$ gauged $O(4)$ Skyrme system on $\R^3$. 
However, these models possess generalizations for other dimensions of the background geometry,
solitons of the $SO(2)$ gauged $O(3)$ Skyrme system on $\R^2$ being constructed
by Schoers~\cite{Schroers:1995he}.
The simpler problem of gauging the planar Skyrmions~\cite{Schroers:1995he} 
is much more transparent, and has led to a proposal for the
$SO(D)$ gauging of $O(D+1)$ Skyrme system on $\R^D$, in Ref.~\cite{Tchrakian:1997sj}.

A lower bound on the energy of a gauged Skyrmion in $D$ dimensions persists also for $SO(N)$ gauge groups with $2\le N\le D$, as $e.g.$, in \cite{Callan:1983nx,Piette:1997ny,Radu:2005jp}. 
The prescription for constructing such lower bounds is systematically explained in Appendix {\bf A} 
of \cite{Tchrakian:2015pka}, where, 
in the specific case of the $O(5)$ Skyrme model on $\R^4$ of interest here, 
only one pair of the components of the $5$-component
Skyrme scalar are gauged. This is unsatisfactory in the context of the problem at hand, where it is desirable
to gauge two pairs of the Skyrme scalar with $SO(2)$, with the aim of imposing bi-azimuthal symmetry in $\R^4$.  
Such a gauging prescription together with the corresponding topological charge density is constructed in Appendix {\bf A} of the present paper.
There we start
with the density pertaining to the system gauged with the full $SO(4)$ group,
which acts on four of the components of the $5$-component Skyrme scalar, and then perform  a group contraction to $SO(2)$.
Here, and in Refs.~\cite{Tchrakian:1997sj,Tchrakian:2015pka} the integral of this lower bound is loosely described as a ``topological charge''
in analogy with its Higgs analogue.

 \medskip   

Systematic and quantitative studies of $SO(2)$ gauged Skyrme systems in $2+1$ dimensions 
were recently carried out in \cite{Navarro-Lerida:2016omj,Navarro-Lerida:2018giv,Navarro-Lerida:2018siw}.
There, the emphasis was on the effect of the  Chern-Simons dynamics,
which is possible to define in all odd spacetime dimensions~\footnote{The prescription of constructing Chern-Simons densities for gauged Skyrme systems in
even spacetime dimensions is given in \cite{Tchrakian:2015pka}, but to date its effect has not been quantitatively studied.}. It was found that the presence of the Chern-Simons term resulted in a
{\it non standard} relation between mass and charge/spin, seen in \cite{Navarro-Lerida:2016omj}, 
and in the dissipation of the baryon number,
 seen in \cite{Navarro-Lerida:2018giv,Navarro-Lerida:2018siw}. 
Both these effects are striking new results, and it is not unreasonable
to expect that they are not exclusive to this low dimension only.

To carry out such a study in the physically most interesting case of the Abelian gauged Skyrmion in $3+1$ dimensions is technically a very substantial problem since the definition of the
relevant Chern-Simons density, proposed in Ref.~\cite{Tchrakian:2015pka}, involves a  supplementary $O(6)$ Skyrme scalar in addition to the  $O(4)$ Skyrme scalar that describes the nucleon.
It is therefore reasonable to defer that investigation and proceed instead with the study of an Abelian gauged $O(5)$ Skyrmion in $4+1$ (odd) dimensions, where a (usual) Chern-Simons density is defined.
This is the remit of the present work, which is a preliminary attempt at inquiring whether the phenomena exposed in Refs.~\cite{Navarro-Lerida:2016omj,Navarro-Lerida:2018giv,Navarro-Lerida:2018siw}
persist.

In $3+1$ dimensions, irrespective of the absence of the (usual) Chern-Simons density there is the additional technical complication that the largest symmetry that can be applied to the Abelian field 
is axial symmetry, resulting in a two-dimensional residual subsystem~\cite{Piette:1997ny,Radu:2005jp,Navarro-Lerida:2018giv}. This is of course the case in all higher even dimensional spacetimes, which
results in the necessity of tackling multidimensional partial differential
equations (PDEs) of the reduced systems. An advantage of $4+1$ dimensions, $i.e.$, static $\R^4$, is that it is possible to
impose an enhanced symmetry on the system that renders the residual system one-dimensional, depending only on the radial variable.
The enhanced symmetry in question is that which is imposed on the the bi-polar (bi-azimuthal) symmetry in $\R^4$, and leads to a simplified form
of the equations which are solved in a numerical approach.
 
\medskip   

In the present work we study solutions in both the purely magnetic sector where the electric component of the Abelian field vanishes, 
$A_0=0$,
and also when $A_0\neq 0$, where electric charge and angular momentum are present.
 In both sectors, we consider the symmetry enhanced systems
resulting in one dimensional ODEs. In the purely magnetic sector, the fully bi-azimuthal solutions to two dimensional PDEs are also constructed.
In the $A_0\equiv V \neq 0$ sector, where only radial solutions were considered, the electric charge $Q_e$ was given by the
asymptotic solutions for $V=V_0+\frac{Q_e}{ 4\pi^2 r^2}+\dots$, in agreement with the solution of the Laplace equation on $\R^4$. In this sense, our electrically (and magnetically) charged solutions are
analogues of the Julia-Zee dyons~\cite{Julia:1975ff} in $3+1$ dimensions. This
definition of electric charge contrasts with that given by Paul and Khare~\cite{Paul:1986ix}
for the Abelian gauged Maxwell--Chern-Simons system interacting with a scalar in $2+1$ dimensions. While in the latter~\cite{Paul:1986ix} the existence of electric charge and spin depends on a nonvanishing Chern-Simons (CS) density, here these are present independently of CS
dynamics as in the case of JZ dyons~\cite{Julia:1975ff}. This is because $Q_e$ in \cite{Paul:1986ix} is proportional to the $first$ Pontryagin charge (and the spin to the square of the Pontryagin charge) on $\R^2$, while here the corresponding quantity is
the $second$ Pontryagin charge on $\R^4$, for which not to vanish the gauge group must be $SU(2)$ and with the Abelian gauge field at hand it vanishes. We plan to return to this question elsewhere.

\medskip   
The paper is structured as follows.
In Section {\bf 2}, we present the model and subject the system to the symmetries described above, and in Section {\bf 3}, we present the results of the numerical analysis. In Section {\bf 4}, we summarise our results and point out to future developments.
In addition, we have supplied two Appendices. Appendix {\bf A} defines the ``topological charge'' supplying the lower bound of the energy. Such a charge density is provided in \cite{Tchrakian:2015pka}, which is not adequate for the present application since only two of the
five components of the Skyrme scalar are gauged in that case. Here, we need to gauge two pairs of Skyrme scalars to enable the imposition of the enhanced symmetry rendering the bi-azimuthal system a radial one. (Appendix {\bf A} in fact stands on its own as a
supplement to the corresponding result in \cite{Tchrakian:2015pka}.) In Appendix {\bf B}, we have established the Belavin inequalities that give the Bogomol'nyi lower bounds, a task which is appreciably more involved than the corresponding one for the ungauged $O(5)$
model, studied in \cite{Brihaye:2017wqa}.

\bigskip
\noindent
 {\bf Conventions}
\\

Throughout the paper, mid alphabet Greek indices, $\mu,\nu,\dots$, label spacetime coordinates running from 1 to 5 (with $x^5=t$). When referring to spacelike coordinates only we will use mid alphabet Latin letters, $i,j,\dots$. Early Latin letters, $a,b,\dots $ label the internal indices of the scalar field multiplet, running from 1 to 5, when primed $a',b',\dots $ they just refer to the first four internal indices 1, 2, 3, 4. Since we will gauge the Skyrme scalar field by pairs of components, we will indicate the pair $(1,2)$ by early Greek letters, $\alpha,\beta,\dots$, while for the pair $(3,4)$ we will employ early capital Latin indices, $A,B,\dots$.
As standard, we use Einstein's summation convention, but to alleviate  notation, no distinction is made between covariant and contravariant \textit{internal} indices.

The background of the theory is Minkowski spacetime, where the spatial $\mathbb{R}^4$ is written in terms of bi-polar spherical coordinates,
\begin{eqnarray}
\label{Mink}
ds^2=dr^2+r^2d\Omega_3^2-dt^2~,
\end{eqnarray}
 where $r,t$ are the radial and time coordinates respectively, 
while $d\Omega_3^2$ is the metric of the three sphere, with
\begin{eqnarray}
 d\Omega_3^2=d\theta^2+\sin^2 \theta d\varphi_1^2+\cos^2 \theta d\varphi_2^2,
\end{eqnarray} 
where
$0\leq \theta\leq \pi/2$
and
$0\leq \varphi_{1,2}<2\pi$.

In addition to using the coordinates $(r,\ta)$, we will find it convenient to employ instead
\be
 \rho=r\sin\ta\ ,\quad\si=r\cos\ta\,,
\ee
(with $0\leq \sigma,\rho<\infty$)
in some expressions, such that (\ref{Mink}) becomes
\begin{eqnarray}
\label{Mink1}
ds^2=d\rho^2+\rho^2 d\varphi_1^2+d\sigma^2 +\sigma^2   d\varphi_2^2-dt^2.
\end{eqnarray}

\section{The model }

\subsection{Gauging prescription and the action }
In $4+1$ spacetime dimensions, the Skyrme model  
 is described by the real scalar field $\f^a=(\f^{\al},\f^{A},\f^5)$, $\al=1,2\ ;\ A=3,4$, 
subject to the sigma model constraint
\begin{eqnarray}
|\f^a|^2=(\f^{\al})^2+(\f^{A})^2+(\f^5)^2=1\,. \label{constraint}
\end{eqnarray}

The gauging prescription~\cite{Tchrakian:1997sj}-\cite{Tchrakian:2015pka} for a $O(D+1)$ 
Skyrme scalar in $D+1$ spacetime involves gauging only the first $D$ components, 
$\f^{a}\ ,\ \ a=1,\dots,D$. 
Thus in the present case in $4+1$ dimensions, only the first
four components $\f^{a}\ ,\ a=1,2,3,4$ are gauged. 
The gauging prescription in the present case is stated by the definition of the covariant derivatives
\bea
\f_\mu^{\al}&=&D_\mu\f^{\al}=\pa_\mu\f^{\al}+A_\mu(\vep\f)^{\al} \ ,
\label{coval}\\
\f_\mu^{A}&=& D_\mu\f^{A}=\pa_\mu\f^{A}+A_\mu(\vep\f)^{A} \ ,
\label{covA}
\\                  
\f_\mu^{5}&=&D_\mu\f^5=\pa_\mu\f^5
\label{cov3}\, ,
\eea
with $\al=1,2$; $A=3,4$. Here $\vep$ denotes the Levi-Civita symbol in each of the two-dimensional subsets of internal indices, $(1,2)$ and $(3,4)$, respectively. More specifically, $(\vep\f)^1 = \f^2 , (\vep\f)^2 = -\f^1 $ and similar for indices $(3,4)$.

In what follows, we shall use the abbreviated notation
\[
\f_{\mu_1\mu_2\dots\mu_p}^{a_1a_2\dots a_p}(p)=\f_{\mu_1}^{a_1}\wedge\f_{\mu_2}^{a_2}\dots\wedge\f_{\mu_p}^{a_p} \ ,
\]
for the $p$-fold antisymmetrised products of the $1$-form $\f_\mu^a$ defined by \re{coval}-\re{cov3}. The squares of these quantities  describe the Skyrme kinetic terms,
which in this case are allowed for $p=1,2,3,4$. Here, we restrict our attention to the quadratic, quartic and sextic
terms with $p=1,2,3$, and eschew the octic term with $p=4$.

We will use the further abbreviated notation
%
%
\begin{eqnarray}
&&
\nonumber
\F^2\equiv \phi^a_{\mu_1} \phi^{a}_{\mu_2} g^{\mu_1 \mu_2}\ ,
\\
\label{calF}
&&
\F^4\equiv \phi^{ab}_{\mu_1 \nu_1} \phi^{ab}_{\mu_2 \nu_2} g^{\mu_1 \mu_2}g^{\nu_1 \nu_2} \ ,
\\
&&
\nonumber
\F^6 \equiv \phi^{abc}_{\mu_1 \nu_1\la_1} \phi^{abc}_{\mu_2 \nu_2\la_2} g^{\mu_1 \mu_2} g^{\nu_1 \nu_2}g^{\la_1 \la_2} \ ,
\end{eqnarray}
where $g^{\mu\nu}$ are contravariant to the metric tensor $g_{\mu\nu}$ of the five dimensional background geometry. Thus,
$\phi^{\mu}_a=\phi^{a}_\nu g^{\mu\nu}$, $\phi^{\mu\nu}_{ab}=\phi^{ab}_{\rho\si}g^{\mu\rho}g^{\nu\si}$,  $etc.$.
Note that $a,b,\dots$ are the global $O(5)$ indices for which we do not distinguish $upper$ and $lower$, for typographical convenience.

We consider the following action
\begin{eqnarray}
\label{action}
S= \int d^5x \sqrt{-g}\,
 \left [
 \frac{\lambda_1}{2}\F^2+\frac{\lambda_2 }{4}\F^4
+\frac{\lambda_3}{36} \F^6+\lambda_0 U (\phi^5)
+
 \frac{1}{4} \lambda_M F_{\mu\nu} F^{\mu\nu}
+\frac{\ka}{\sqrt{-g}}
\vep^{\mu\nu\rho\si\la}A_\la F_{\mu\nu}F_{\rho\si}
 \right ],
\end{eqnarray}
which apart from the above quantities features the (standard) Maxwell and Chern-Simons terms, and  a Skyrme potential $U$. 
Thus
$F_{\mu\nu}$ is the Maxwell field $F_{\mu\nu}= \partial_\mu A_\nu - \partial_\nu A_\mu$, $A_\mu$ being the gauge connection. Our choice for the Skyrme potential is
\be
U=1-\f^5 \ ,\label{Skpot}
\ee
which is the analogue of the ``pion mass potential'', often used in the three dimensional and planar Skyrme models.
$(\la_0,\la_1,\la_2,\la_3,\lambda_M)\geqslant 0$ are coupling constants.
We shall also define
\begin{eqnarray}
\lambda_{M}=\frac{1}{g^2},
\end{eqnarray}
with $g$ the gauge coupling constant, such that the gauge decoupling limit in \cite{Brihaye:2017wqa} is approached for $g=0$.


Again, to accommodate the eventual formulation in curved coordinates, we replace all partial derivatives $\partial_\mu$ in \re{coval}-\re{cov3} formally by $\nabla_\mu$.
Varying the Lagrangian \re{action} $w.r.t.$ the scalars $\phi^a$ leads to the Euler-Lagrange equations
\be
\label{EL1}
\left(\delta^{d a}-\f^{d }\f^{a}\right)
\left\{
2\la_1\, D^{\mu}\f_{\mu}^{a}
+8\la_2\,\f^{\nu}_{b} D^{\mu}\f_{\mu\nu}^{ab}
+9\la_3\,\f_{bc}^{\nu\la} D^{\mu}\f_{\mu\nu\la}^{abc}
+\la_0\,\frac{\pa U}{\pa\f^{a}}
\right\}=0 \ ,
\ee
while the corresponding equations for the Maxwell field are
\begin{eqnarray}
\label{Max}
\lambda_M \nabla_\nu F^{\nu \mu}=J^\mu+\kappa \vep^{\mu\nu\rho\si\la}  F_{\nu\rho}F_{\si\la},
\end{eqnarray}
where $J^\mu=J^\mu[\f(p)]$ is the Skyrme current arising from the variation 
$w.r.t.$ the Maxwell potential.

Variation of (\ref{action}) $w.r.t.$ the metric tensor $g_{\mu\nu}$ leads to the energy-momentum tensor of the model
\begin{eqnarray}
 T_{\mu\nu} = \lambda_M T_{\mu\nu}^{(M)}+\lambda_0 T_{\mu\nu}^{(0)}+\lambda_1 T_{\mu\nu}^{(1)}+\lambda_2 T_{\mu\nu}^{(2)}+\lambda_3 T_{\mu\nu}^{(3)} \ ,
\end{eqnarray}
in terms of the contributions of the distinct terms in (\ref{action}), which read
\begin{eqnarray}
\nonumber
&&
 T_{\mu\nu}^{(M)} = F_{\mu\rho}F_{\nu\si}g^{\rho\si}-\frac{1}{4}g_{\mu\nu}F_{\rho\si}F^{\rho\si},
\\
\nonumber
&&
 T_{\mu\nu}^{(0)} =-g_{\mu\nu} U (\phi^5),
\\
\nonumber
&&
 T_{\mu\nu}^{(1)} = \phi^a_{\mu} \phi^a_{\nu}-\frac{1}{2}g_{\mu\nu}\F^2,
\\
\label{tij}
&&
 T_{\mu\nu}^{(2)} =   \phi^{ab}_{\mu\rho} \phi^{ab}_{\nu\si}g^{\rho\si}-\frac{1}{4}g_{\rho\si}\F^4 ,
\\
\nonumber
&&
 T_{\mu\nu}^{(3)} = \frac{1}{6}\left(\phi^{abc}_{\mu\rho\tau} \phi^{abc}_{\nu\si\la}g^{\rho\si} g^{\tau\la}-\frac{1}{6}g_{\mu\nu}\F^6 \right).
\end{eqnarray}
%

As usual, the  $tt$ component of the mixed energy-momentum tensor, ${T^t}_t$, 
(taken with minus sign)
 corres\-ponds to the local mass-energy density,
while the angular momentum densities in the $(x_1,x_2)$ and $(x_3,x_4)$ planes are given by the $t \varphi_1$ and $t \varphi_2$ components, ${T^t}_{\varphi_1}$ and ${T^t}_{\varphi_2}$, respectively.

\subsection{The Ansatz and boundary conditions }
\subsubsection{The general case }
We consider a U(1) Ansatz in terms of three potentials, two magnetic
$a_{1,2}(r,\theta)$, and one electric, $V(r,\theta)$, with
\begin{eqnarray}
\label{gauge-ansatz}
A=a_{1}(r,\theta) d\varphi_1+a_{2}(r,\theta) d\varphi_2+V(r,\theta) dt,
\end{eqnarray} 
and the following expression of the scalars
\begin{eqnarray}
\nonumber
&&
 \phi^1=\Psi_1(r,\theta) \cos (n_1 \vf_1- \om t) ,~~\phi^2=\Psi_1(r,\theta) \sin (n_1 \vf_1 -\om t) ,
\\
&&
\label{Skyrme-ansatz}
\nonumber
\phi^3=\Psi_2(r,\theta) \cos (n_2 \vf_2-\om t) ,~~\phi^4=\Psi_2(r,\theta) \sin (n_2 \vf_2-\om t) ,
\\
&&
\phi^5=\Psi_3(r,\theta)~,
\end{eqnarray} 
with $n_1,n_2$ two positive integers (the winding numbers) 
and $\om \geq 0$ the field frequency.
Also, the functions $\Psi_1, \Psi_2, \Psi_3$ are subject to the
constraint (\ref{constraint})
\begin{eqnarray}
\label{constr}
 \Psi_1^2 + \Psi_2^2+ \Psi_3^2=1.
\end{eqnarray} 

In this approach, the problem reduces to solving a set of six PDEs
  with dependence on only two coordinates.
As usual, these equations result by varying (\ref{action}) $w.r.t.$ the functions $ \Psi_1, \Psi_2, \Psi_3$
and $ a_1,a_2,V$, respectively.
Before stating the boundary conditions, we display the expression of several 
terms which enter the action $S$ (\ref{action}),
 subject to the general Ans\"atze \re{gauge-ansatz}-\re{Skyrme-ansatz}. 
\begin{eqnarray}
\frac{1}{2}F_{\mu \nu}F^{\mu \nu}=\frac{1}{r^2}
\left[
\frac{1}{\sin^2\theta} \left(a_{1,r}^2+\frac{1}{r^2}a_{1,\theta}^2 \right)
+\frac{1}{\cos^2\theta} \left (a_{2,r}^2+\frac{1}{r^2}a_{2,\theta}^2\right)
-\left (V_{,r}^2+\frac{1}{r^2}V_{,\theta}^2  \right)
\right]~,
\end{eqnarray}
\begin{eqnarray}
\vep^{\mu\nu\rho\si\la}A_\la F_{\mu\nu}F_{\rho\si}=
\frac{8}{\sin\theta \cos\theta}
\left[
\left(a_1 a_{2,\theta} -a_2 a_{1,\theta} \right)V_{,r}
+\left(a_2 a_{1,r} -a_1 a_{2,r} \right) V_{,\theta}
+\left( a_{2,r} a_{1,\theta} - a_{1,r} a_{2,\theta}  \right)V
\right],
\end{eqnarray}
and
\begin{eqnarray}
&&\F^2\equiv \phi^a_{i_1} \phi^{a}_{i_2} g^{i_1 i_2}=
\Psi^2_{1,r}+ \Psi^2_{2,r}+ \Psi^2_{3,r}
+\frac{1}{r^2}
\left(
 \Psi^2_{1,\theta}+ \Psi^2_{2,\theta}+ \Psi^2_{3,\theta}
\right)
					\\
	\nonumber
	&&
	{~~~~}
+\frac{1}{r^2}
\left[
          \left(
 \frac{(n_1-a_1)^2}{\sin^2\theta}   +\frac{a_2^2}{\cos^2\theta} 
          \right)\Psi_1^2 
	+          \left(
 \frac{(n_2-a_2)^2}{\cos^2\theta}   +\frac{a_1^2}{\sin^2\theta} 
          \right)\Psi_2^2 
\right]
-(\Psi_1^2 +\Psi_2^2 )(\om+V)^2~,
\end{eqnarray}
while the expressions of 
$\F^4$ and 
$\F^6$ are too complicated to include here.

 The boundary conditions satisfied by the functions $ \Psi_1, \Psi_2, \Psi_3$ are
\begin{eqnarray}
\label{bc1}
&&
\Psi_1\big|_{r=0}=-1,~~\Psi_2\big|_{r=0}=0,~~\Psi_3\big|_{r=0}=-1,
~~\Psi_1\big|_{r=\infty}=0,~~\Psi_2\big|_{r=\infty}=0,~~\Psi_3\big|_{r=\infty}=1,
\\
\nonumber
&&
\Psi_1\big|_{\theta=0} =\partial_\theta \Psi_2\big|_{\theta=0}=\partial_\theta \Psi_3\big|_{\theta=0}=0,~~
\partial_\theta \Psi_1\big|_{\theta=\pi/2} =  \Psi_2\big|_{\theta=\pi/2}=\partial_\theta \Psi_3\big|_{\theta=\pi/2}=0,
\end{eqnarray}
while for the gauge potentials we impose
\begin{eqnarray}
\label{bc2}
&&
a_1\big|_{r=0}=~a_2\big|_{r=0}=0,~~V_{,r}\big|_{r=0}=0,
~~a_1\big|_{r=\infty}=a_2\big|_{r=\infty}=0, ~~V\big|_{r=\infty}=V_0,
\\
\nonumber
&&
a_1\big|_{\theta=0} =\partial_\theta a_2\big|_{\theta=0}=0,~~
\partial_\theta a_1\big|_{\theta=\pi/2} = a_2\big|_{\theta=\pi/2}=0, 
~~\partial_\theta V\big|_{\theta=0,\pi/2}=0.
\end{eqnarray}
These boundary conditions are compatible with an approximate form of the solutions on the boundaries of the domain of integration, together with some physical requirements ($e.g.$, regularity and finiteness of global charges). Another criteria here (and an important guideline in selecting among possible sets of boundary conditions)  is the compatibility with the spherically symmetric ungauged limit in \cite{Brihaye:2017wqa}, together with the radially enhanced limit in Section 2.2.2. For example, Eqs. (\ref{n11}) and (\ref{n12}) in Section 2.2.2 imply that the functions $\Psi_1$ and $a_1$ vanish at $\theta=0$, while $\Psi_2$, $\Psi_3$, $a_2$ and $V$ should satisfy Newman boundary conditions. Then one assumes the existence of a generic small $\theta$-expansion of the form ${\cal U}=\sum_{k\geq 0}u_k(r) \theta^k$ (with ${\cal U}=\{ \Psi_1,\Psi_2,\Psi_3;a_1,a_2,V\}$) which is plugged into the field equations. (Note that the coefficients $u_0(r)$ vanish for $\Psi_1$ and $a_1$, while $u_1(r)$ is zero for the remaining functions). A similar approach is implemented for $\theta=\pi/2$ and at the limits of the $r$-interval.

As usual the total mass-energy, $M$, and angular momenta, $J_{1,2}$, of a solution are defined as 
\begin{equation}
\label{MJ}
M=-\int d^4 x \sqrt{-g}{T^t}_t,~~
J_1=  \int d^4 x \sqrt{-g}{T^t}_{\varphi_{1}},~~
J_2=  \int d^4 x \sqrt{-g}{T^t}_{\varphi_{2}}~,
\end{equation}
while the electric charge $Q_e$ is computed from the electric flux at infinity,
\begin{equation}
\label{Qe}
Q_e=\oint_\infty dS_{rt}F^{rt},~
\end{equation}
and thus
can also be evaluated from the asymptotics  of the electric  potential
\begin{equation}
\label{Qe1}
V=V_0+\frac{Q_e}{ 4\pi^2 r^2}+\dots~,
\end{equation}
with $V_0$ a constant.
However, by using the field equations, the volume integral in the expression (\ref{MJ})
of $J_{1,2}$
 can be converted into  
surface integrals at infinity in terms of Maxwell potentials, and one finds\footnote{Note, however,
that the corresponding densities are not equal.}
\begin{equation}
\label{JQ} 
J_1= \frac{1}{2} \lambda_M n_1  Q_e,~~
J_2=  \frac{1}{2} \lambda_M  n_2 Q_e.
\end{equation}

\subsubsection{$n_1=n_2=1$: a symmetry enhanced Ansatz}
Remarkably, it turns out that the choice
\begin{eqnarray}
\label{n11}
&&
\Psi_1=\sin \psi(r) \sin \theta,~~\Psi_2= \sin \psi(r) \cos \theta, ~~\Psi_3=\cos \psi(r)  ,
\\
\label{n12}
&&
a_{1}(r,\theta)=a_{\varphi}(r)\sin^2 \theta,~~a_{2}(r,\theta)=a_{\varphi}(r)\cos^2 \theta,~~V(r,\theta)=V(r)~,
\end{eqnarray} 
provides a consistent factorization of the angular dependence for 
the general model, provided that\footnote{This is similar to the factorization of the $\theta-$dependence on the $S^3$-sphere  
employed in the scalar field Ans\"atze in 
\cite{Hartmann:2010pm},
\cite{Dias:2011at}.}
\begin{eqnarray}
\label{new-ans1}
n_1=n_2=1.
\end{eqnarray} 
This restrictive Ansatz greatly reduces the complexity of the system 
and simplifies the numerical construction of  the lowest topological charge solutions,
which are found in this case by solving a set of three ordinary differential equations (ODEs).
For example,
with the above Ansatz,  the  effective action
of the model reads
 \begin{eqnarray}
L_{\rm eff}&=&r^3
                    \left\{
\frac{\lambda_M }{2r^2}
\left(
 a_{\varphi}'^2+\frac{4 a_\varphi^2}{r^2}-r^2V'^2
\right)+\frac{16\ka}{r^3}    \left( V a_{\varphi}'-a_{\varphi} V'  \right)a_{\varphi}
\right.
\nonumber
\\
&&+\frac{\lambda_1}{2}
\left[
\psi'^2+\frac{\sin^2 \psi}{r^2}(2+(1-a_\varphi)^2 -r^2(\om+V)^2 )
\right]
\\
\nonumber
&&
+\lambda_2
\frac{\sin^2 \psi}{r^2}
\left[
\psi'^2 \left(2+(1-a_\varphi)^2-r^2 (\om+V)^2 \right) +\frac{\sin^2 \psi}{r^2} \left(1+2 ((1-a_\varphi)^2 -r^2(\om+V)^2) \right)
\right]
\\
\nonumber
&&
\left.   +\lambda_3
\frac{\sin^4\psi}{r^4}
\left[
\psi'^2  \left(1+2(1-a_\varphi)^2-2 r^2(\om+V)^2 \right)
+\frac{\sin^2 \psi}{r^2}( (1-a_\varphi)^2-r^2(\om+V)^2 )
\right]
+\lambda_0(1-\cos \psi)
                    \right\},
 \end{eqnarray} 
  the contribution of various terms being transparent.
	The boundary conditions satisfied by the functions 
	$\{\psi(r)$, $a_\varphi(r)$, $V(r)\}$
	results directly from (\ref{bc1})-(\ref{bc2}). 
Also, in this case it is possible to compute an 
approximate form of the solutions at the limits for the domain of integration.
For example, one finds the following small-$r$ expression
 \begin{eqnarray}
 a_\varphi(r)=m_2 r^2+m_4 r^4+O(r^6),~~\psi(r)=\pi+f_1 r+ O(r^3),
~~V(r)=v_0+ v_4 r^4+O(r^5),
 \end{eqnarray} 
 which contains three essential parameters $f_1$, $m_2$ and $v_0$,
while
\begin{eqnarray}
&&
m_4= \frac{96 \kappa^2 m_2^3 }{\lambda_M^2}-\frac{f_1^2(\lambda_1+6f_1^2(\lambda_2+f_1^2\lambda_3))}{12\lambda_M},
\\
\nonumber 
&&
v_4=\frac{f_1^2 (v_0+\om)}{24 \lambda_M}
(\lambda_1+6f_1^2(\lambda_2+f_1^2\lambda_3))
+\frac{\kappa m_2}{\lambda_M^3}[-1152 \kappa^2m_2^3
+f_1^2\lambda_M(\lambda_1+6f_1^2(\lambda_2+f_1^2\lambda_3))].
\end{eqnarray} 
The leading order terms
in the large$-r$ expansion of the solutions are
\begin{eqnarray}
 a_\varphi(r)= \frac{\bar m_2}{ r^2}+\dots,
~~\psi(r)= c\sqrt{\frac{\pi}{2}}\frac{ e^{-r\sqrt{\frac{\lambda_0}{\lambda_1} }}}{r^{3/2}}+\dots,
~~V=V_0+\frac{Q_e}{ 4\pi^2 r^2}+\dots,
 \end{eqnarray} 
with 
$
\bar m_2
$
$Q_e$, $V_0$
and $c$ some constants which are determined by numerics.

The corresponding expressions of the mass-energy and angular momenta
densities are also of interest,
with\footnote{Note that these expressions are given in gauge with $\om=0$, which was employed in numerics.}
 \begin{eqnarray}
&&-{T^t}_t= 
\frac{\lambda_M }{2r^2}
\left(
 a_{\varphi}'^2+\frac{4 a_\varphi^2}{r^2}+r^2 V'^2
\right) +\frac{\lambda_1}{2}
\left[
\psi'^2+\frac{\sin^2 \psi}{r^2}(2+(1-a_\varphi)^2+r^2 V^2)
\right]
\\
\nonumber
&&{~~~~~~~~}
+\lambda_2
\frac{\sin^2 \psi}{r^2}
\left[
\psi'^2 \left(2+(1-a_\varphi)^2+r^2  V^2 \right) +\frac{\sin^2 \psi}{r^2} \left(1+2 ((1-a_\varphi)^2 +r^2 V^2) \right)
\right]
\\
\nonumber
&&{~~~~~~~~}
+\lambda_3
\frac{\sin^4\psi}{r^4}
\left[
\psi'^2  \left(1+2(1-a_\varphi)^2+2 r^2V^2 \right)
+\frac{\sin^2 \psi}{r^2}( (1-a_\varphi)^2+r^2V^2 )
\right]
+\lambda_0(1-\cos \psi) ,
 \end{eqnarray} 
 \begin{eqnarray}
\frac{{T^t}_{\varphi_1}}{\sin^2\theta} =  
\frac{{T^t}_{\varphi_2}}{\cos^2\theta} =-\lambda_M a_\varphi'V'+
\sin^2\psi (1-a_\varphi)V
\left[
\lambda_1+2\lambda_2 \left(\psi'^2+\frac{2\sin^2\psi}{r^2} \right)
+4\lambda_3 \frac{\sin^2\psi}{r^2}
\left(
\psi'^2+\frac{\sin^2\psi}{2r^2}
\right
)
\right]~.
 \end{eqnarray} 
 Then, by using the Maxwell equations,  one can easily show that  
  \begin{eqnarray}
 \sqrt{-g}{T^t}_{\varphi_1}=\sin^3\theta \cos \theta~ {\cal S}',~~  
 \sqrt{-g}{T^t}_{\varphi_2}=\cos^3\theta \sin \theta~ {\cal S}',~~
{\rm with}~~  {\cal S}= \lambda_M (1-a_\varphi)r^3 V' +8 \kappa a_\varphi^2(3-2a_\varphi),
  \end{eqnarray}
which  makes manifest  the
total derivative structure of ${T^t}_{\varphi_i}$.
Also, one observes that despite entering the angular momenta
density,
 the  integral contribution of the CS term vanishes, since $a_\varphi\to 0$
as $r\to 0$ and as $r \to  \infty$.

\subsection{Scaling symmetry and numerical approach}
	
The model  (\ref{action}) 
contains four input parameters $\lambda_i$
together with the gauge coupling constant $g$
(we recall $\lambda_M=1/g^2$).
However, the constant multiplying the 
quadratic term can be taken as an overall  
factor for the Skyrme action.
Also, the equations
are invariant under the transformation 
\begin{eqnarray}
r\to \tau r,~~
 \lambda_0/\lambda_1 \to  \tau^2 \lambda_0/ \lambda_1,~~
 \lambda_2/\lambda_1 \to \lambda_2/ (\tau^2 \lambda_1),~~
 \lambda_3/\lambda_1 \to \lambda_3/ (\tau^4 \lambda_1),
\end{eqnarray}
(with $\tau$ some arbitrary positive parameter),
which was used to
set  $\lambda_3=1$.

Then   the problem still contains 
three free constants $\lambda_0,\lambda_2$ and $\lambda_M$.
In this work, in order to
 simplify the picture,
we have chosen
to solve a model without the quartic term, $i.e.$ we set
\begin{eqnarray}
 \lambda_1=\lambda_3=1,~~\lambda_2=0\ , \label{lambda_choice}
\end{eqnarray} 
such that the only input parameters are $\lambda_0$ and the gauge coupling constant $g$
($i.e.$ the coefficient $\lambda_M$ of the Maxwell term in the action (\ref{action})). The choice (\ref{lambda_choice}) will used for all the numerical solutions presented in this paper.

Starting with the case of solutions within the general Ansatz 
(\ref{gauge-ansatz})-(\ref{Skyrme-ansatz}),
the constraint
(\ref{constr}) is imposed by using the Lagrange multiplier method,
as explained $e.g.$ in \cite{Rajaraman,Radu:2008pp}. 
The numerical calculations were performed by using the professional software 
based on the Newton-Raphson method CADSOL \cite{schoen}. 
The field equations are first discretized on a 
non-equidistant grid and the resulting system is solved iteratively until convergence is achieved.
In this scheme, a new radial variable $x=r/(1+r)$ is introduced which maps 
the semi-infinite region $[0,\infty)$ to the closed region $[0,1]$.
Also, this software package provides error estimates for each unknown function,
which allows judging the quality of the computed solution.
The  numerical error
for the solutions reported in this work is estimated to be typically  $<10^{-4}$. 

The  solutions within the symmetry enhanced Ansatz (\ref{n11})-(\ref{n12})
 were  found   by using the professional
software package COLSYS \cite{COLSYS} 
(although some of them were also computed by using the same approach as in the general case). 
This solver employs a collocation method for boundary-value ODEs
and a damped Newton method of quasi-linearization. 
At each iteration step a linearized problem is solved by using a spline
collocation at Gaussian points.
In this approach, the linearized problem
is solved on a sequence of meshes   until the
successful stopping criterion is reached,
a compactified radial variable $x=r/(1+r)$ being again employed.

\section{Numerical results}

\begin{figure}[ht!]
\begin{center}
\includegraphics[height=.34\textwidth, angle =0 ]{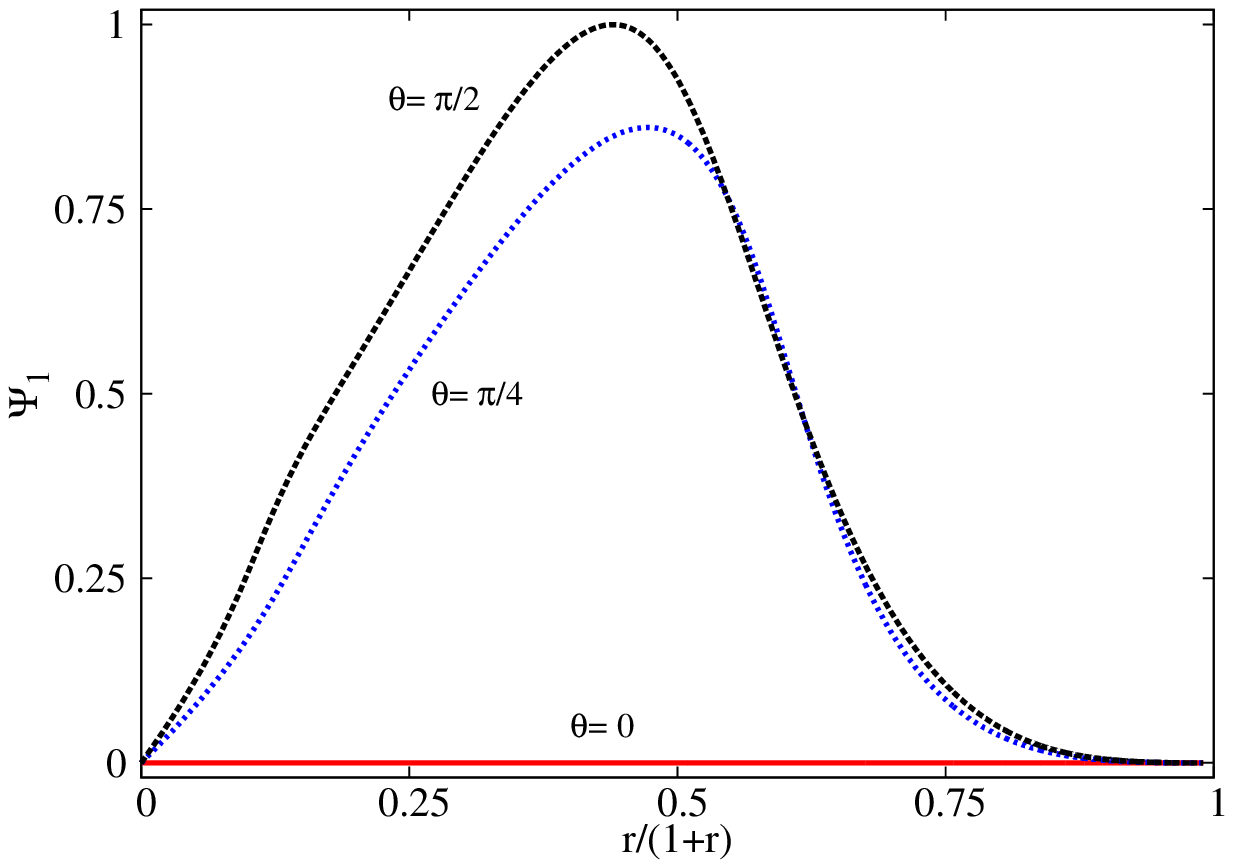} 
\includegraphics[height=.34\textwidth, angle =0 ]{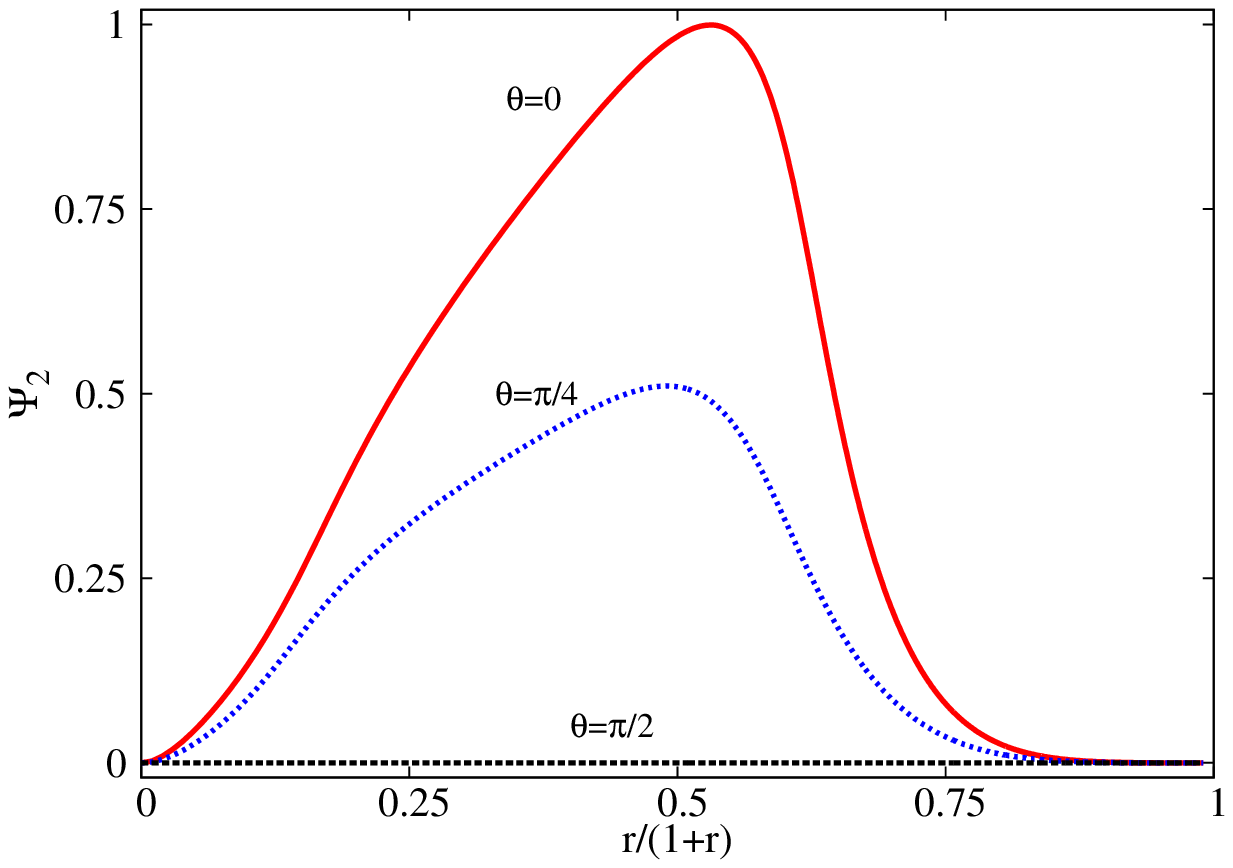}
\includegraphics[height=.34\textwidth, angle =0 ]{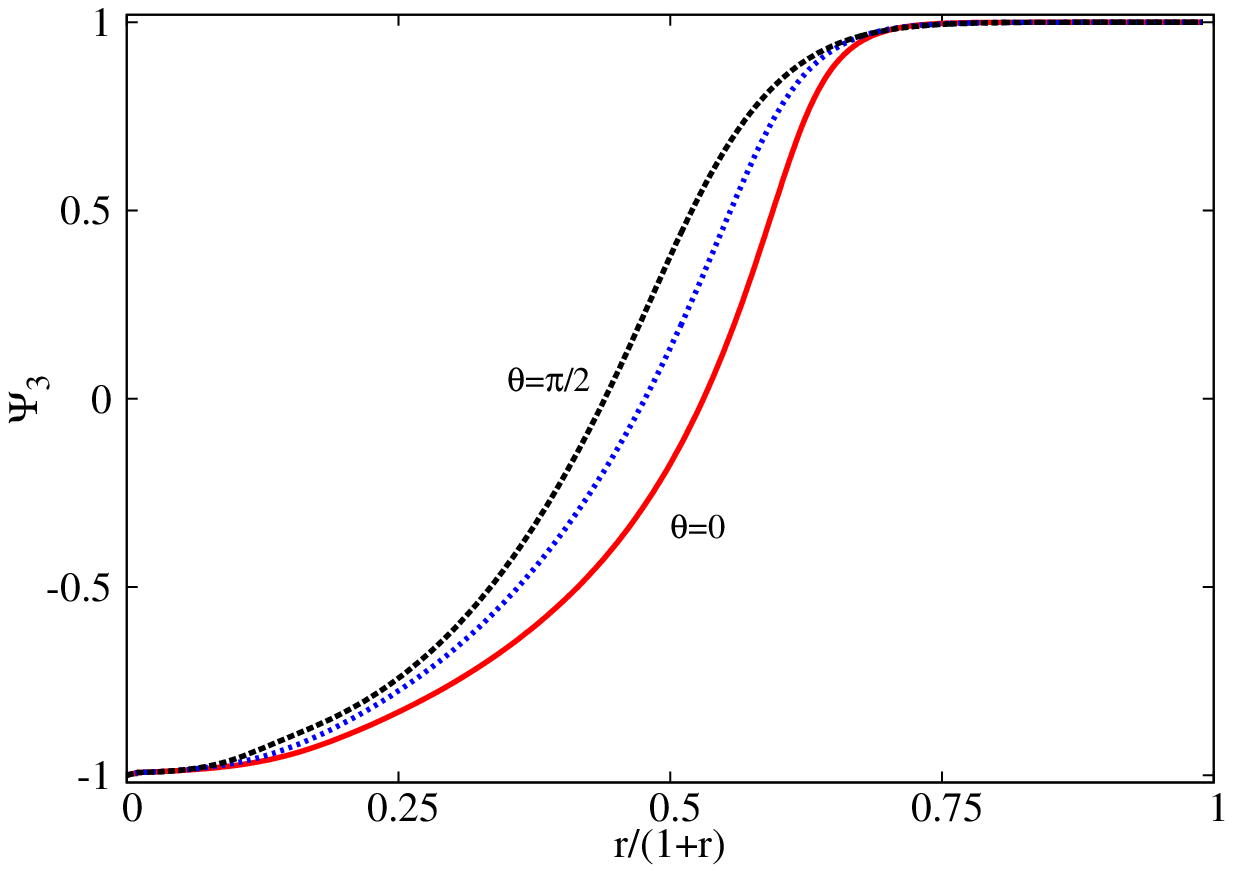} 
\includegraphics[height=.34\textwidth, angle =0 ]{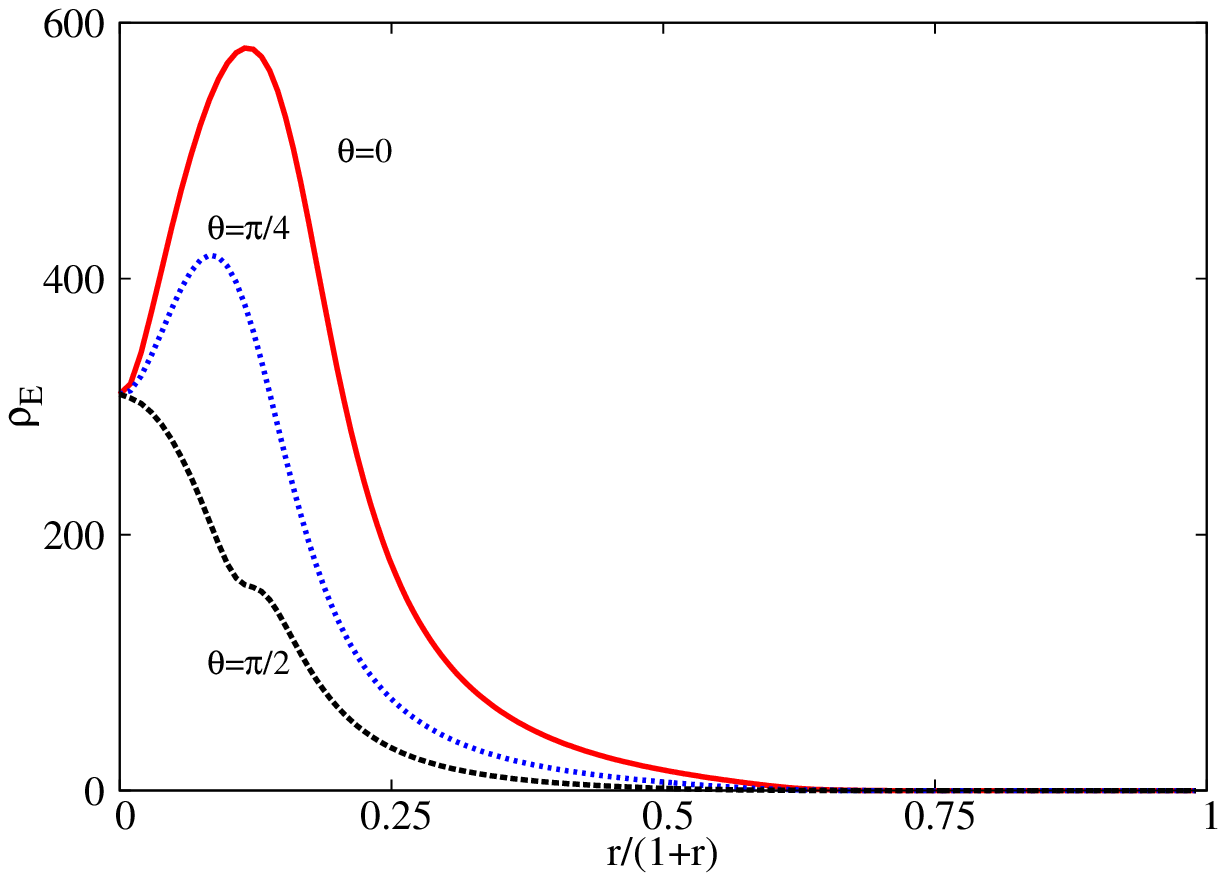}
\includegraphics[height=.34\textwidth, angle =0 ]{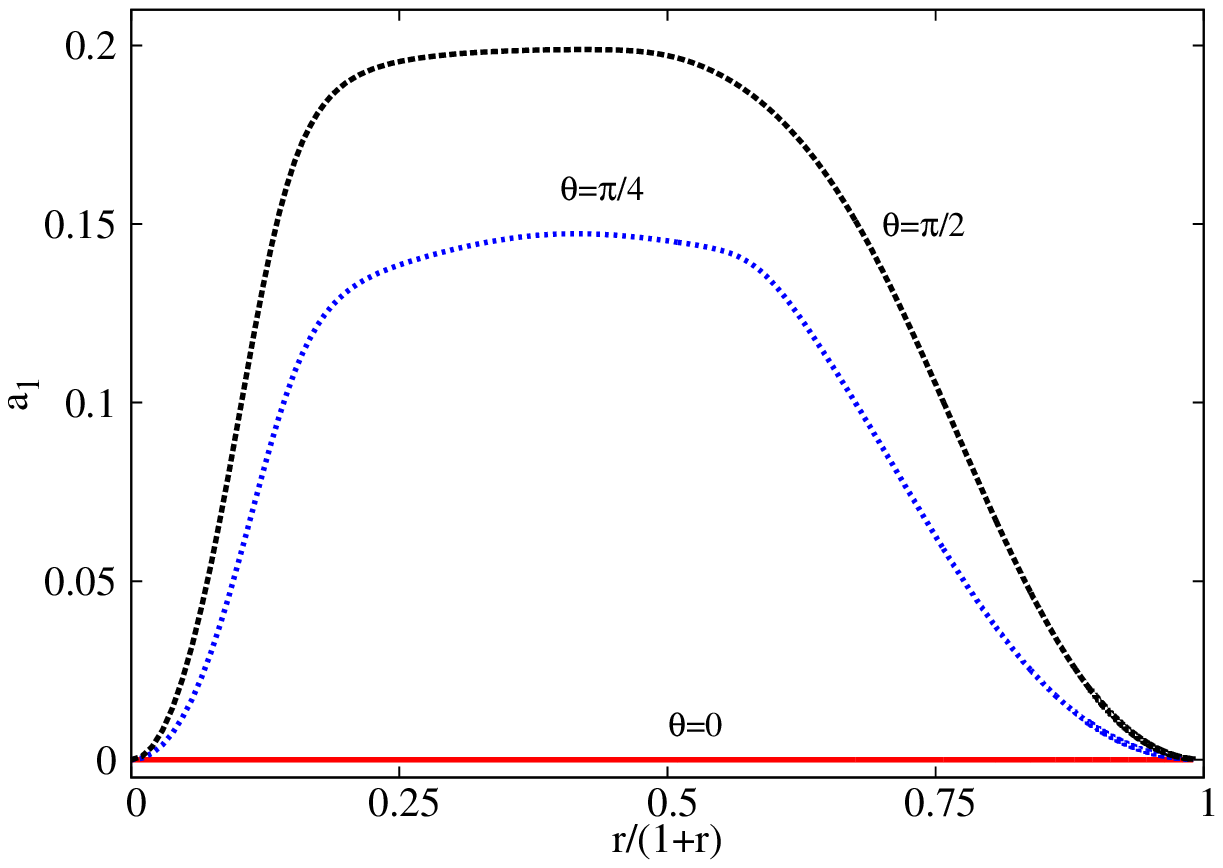} 
\includegraphics[height=.34\textwidth, angle =0 ]{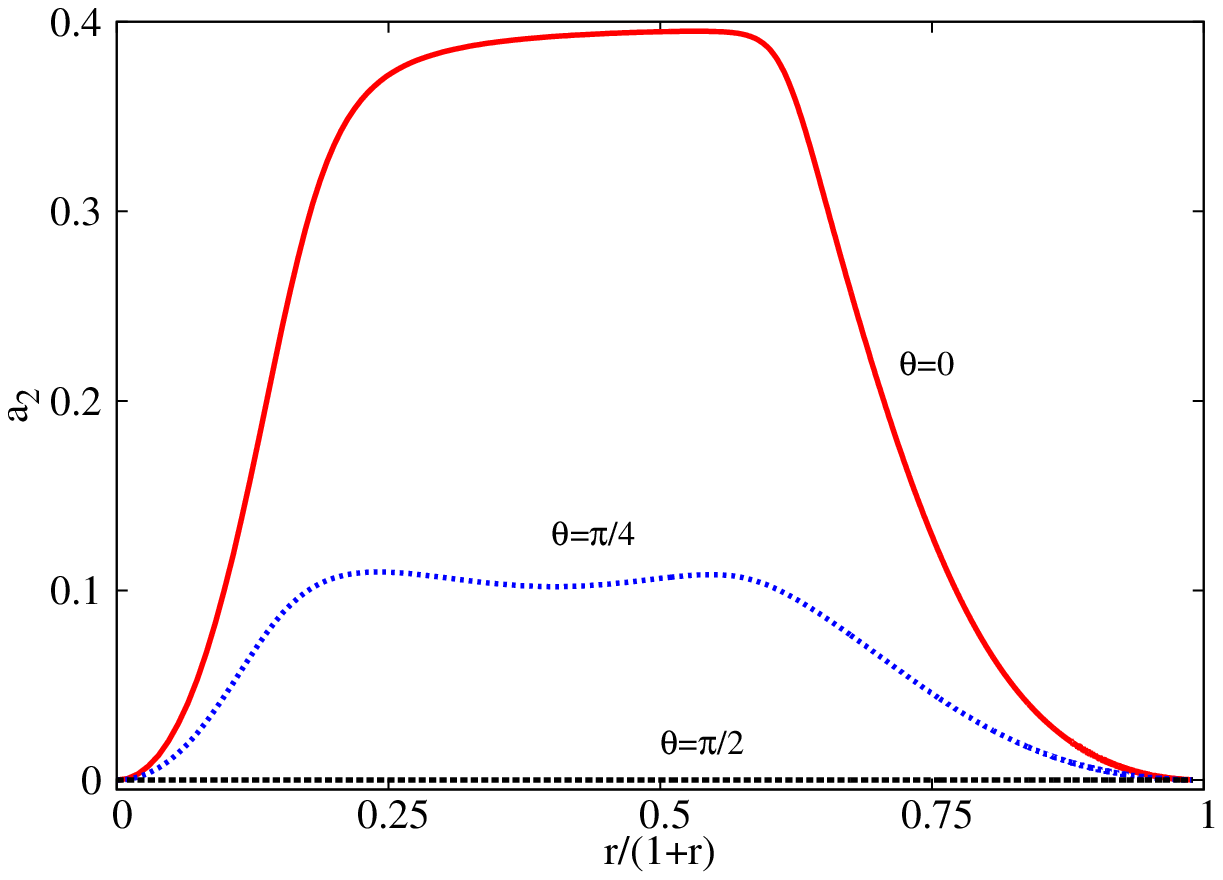}
\end{center}
\caption{
The scalar functions 
$\Psi_1$,
$\Psi_2$,
$\Psi_3$
 and the gauge potentials
$a_1$,
$a_2$ 
which enter the Ansatz 
(\ref{gauge-ansatz})-(\ref{Skyrme-ansatz}) 
are shown together with the mass-energy density $\rho_E=-{T^t}_t$
for a solution with $n_1=1,n_2=2$.
}
\label{plot-2d}
\end{figure}

\subsection{Purely magnetic, static solutions}

\begin{figure}[t!]
\begin{center}
\includegraphics[height=.35\textwidth, angle =0 ]{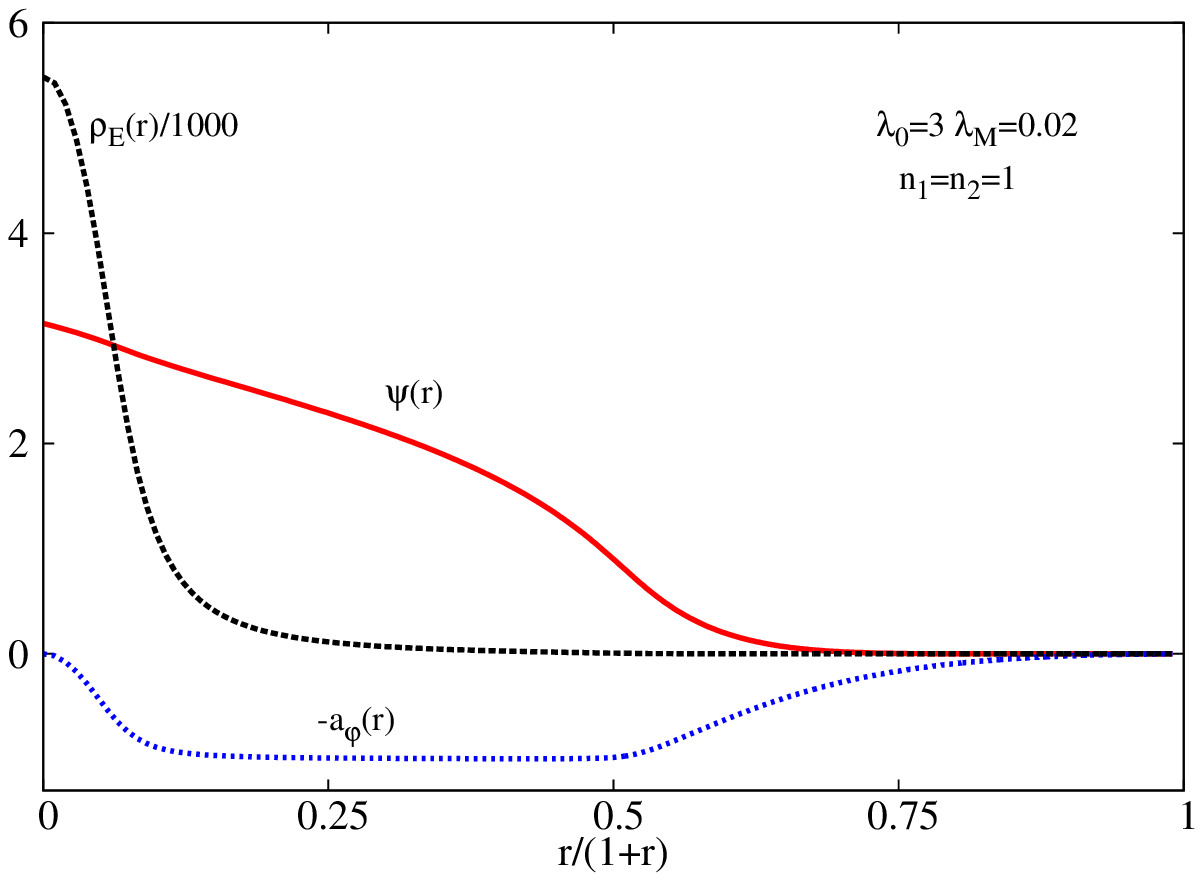} 
\end{center}
\caption{
The scalar function  
$\psi$ 
 and the gauge potential 
$a_\varphi$ 
which enter the simplified  Ansatz 
(\ref{n11})-(\ref{n12}) 
are shown together with the mass-energy density $\rho_E= -{T^t}_t$
for a solution with $n_1=n_2=1$.
}
\label{1d}
\end{figure}

These configurations (characterized by $\kappa=0$) have $V=0$, and their angular momenta and electric
charge density vanish identically.
As an illustrative example for the general case, we show in Figure \ref{plot-2d}
the profiles of a typical solution with $n_1=1,n_2=2$ and  
 $\lambda_M=1/25$, $\lambda_0=0$.
One can see that the gauge and scalar functions (except $\Psi_3$) 
as well as the energy density
 depend strongly on $\theta$.

For completeness, in Figure \ref{1d}
we give a similar plot for 
the special case $n_1=n_2=1$ 
(note however that, here we show the functions $\psi$ and $a_\varphi$ 
which enter the simplified  Ansatz 
(\ref{n11})-(\ref{n12})).

\medskip

\begin{figure}[t!]
\begin{center}
\includegraphics[height=.35\textwidth, angle =0 ]{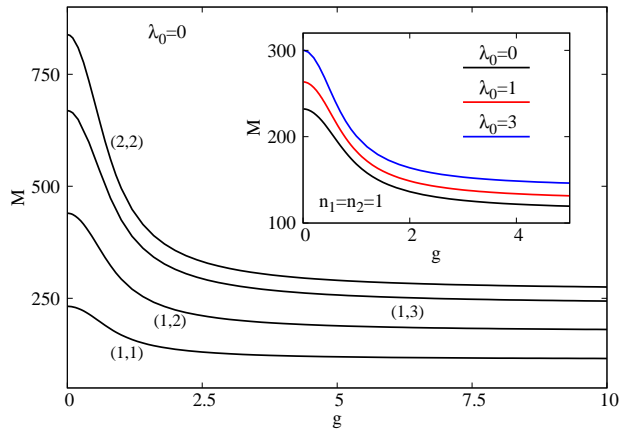}
\end{center}
\caption{
The mass of the $SO(2)$ gauged Skyrmions
is shown as a function of gauge coupling constant $g$ (with $\lambda_M=1/g^2$)
for several values of the winding numbers $n_1,n_2$.
}
\label{Mq}
\end{figure}

The dependence of the solutions on the gauge coupling constant $g$
is shown in Figure \ref{Mq} for the lowest values of the pair $(n_1,n_2)$. One can notice the existence of some universal behaviour there, 
the total mass of the solutions decreasing with $\lambda_M=1/g^2$. The mass is maximized by the  $g=0$ 
configurations, a limit which corresponds  to the ungauged O(5) sigma model, 
whose solutions were discussed in \cite{Brihaye:2017wqa} 
(albeit for the special case $n_1=n_2=1$ only).
A curious result there is that,
for the same other input parameters, the $n_1=n_2=2$ solutions have a larger mass than the  $n_1=1,n_2=3$ configurations.

Although the results there  were found for solutions without a potential, the
same behaviour is found for $\lambda_0\neq 0$, as shown in the inset of Figure \ref{Mq}.
Moreover, as expected, the mass of the solutions always increases with
the parameter
 $\lambda_0$.
Also, one remarks that the generic properties of the static solutions appear to be the same for any choice of the integers $n_1,n_2$.

\begin{figure}[t!]
\begin{center}
\includegraphics[height=.34\textwidth, angle =0 ]{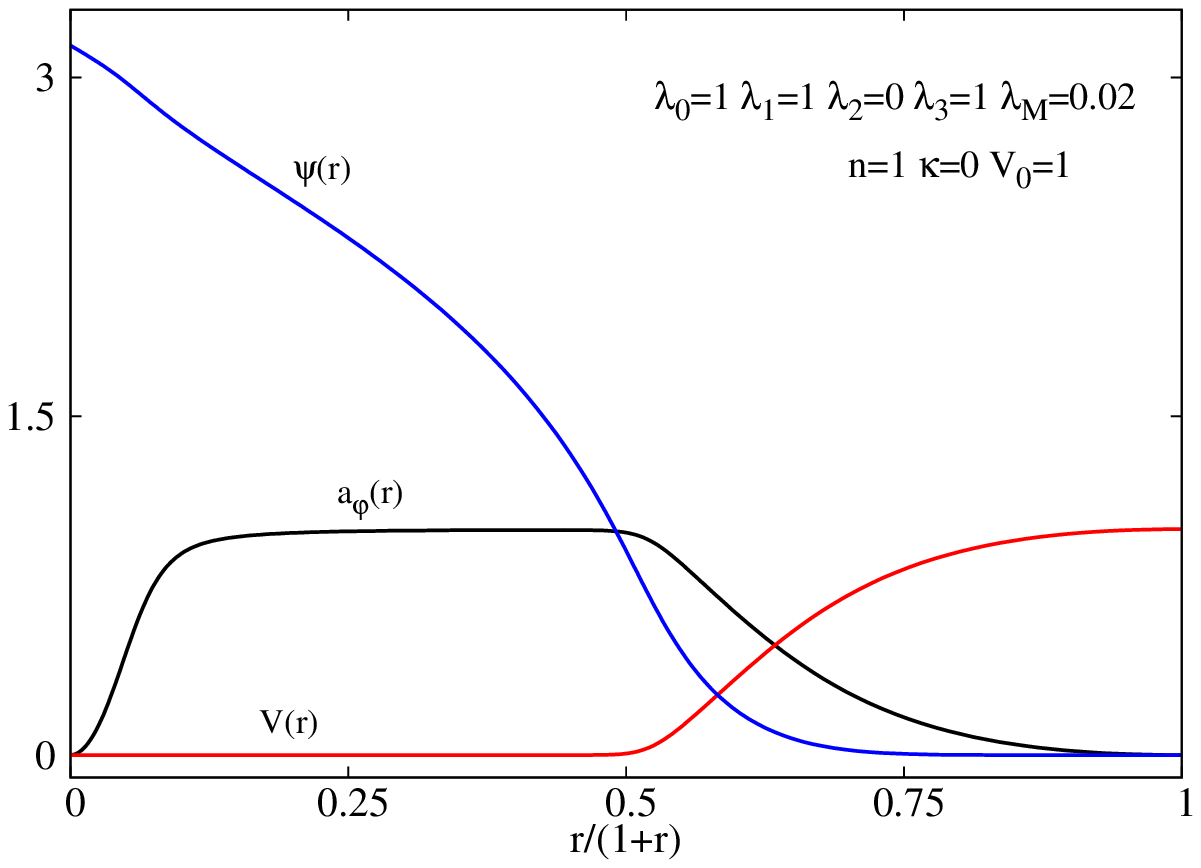}
\includegraphics[height=.34\textwidth, angle =0 ]{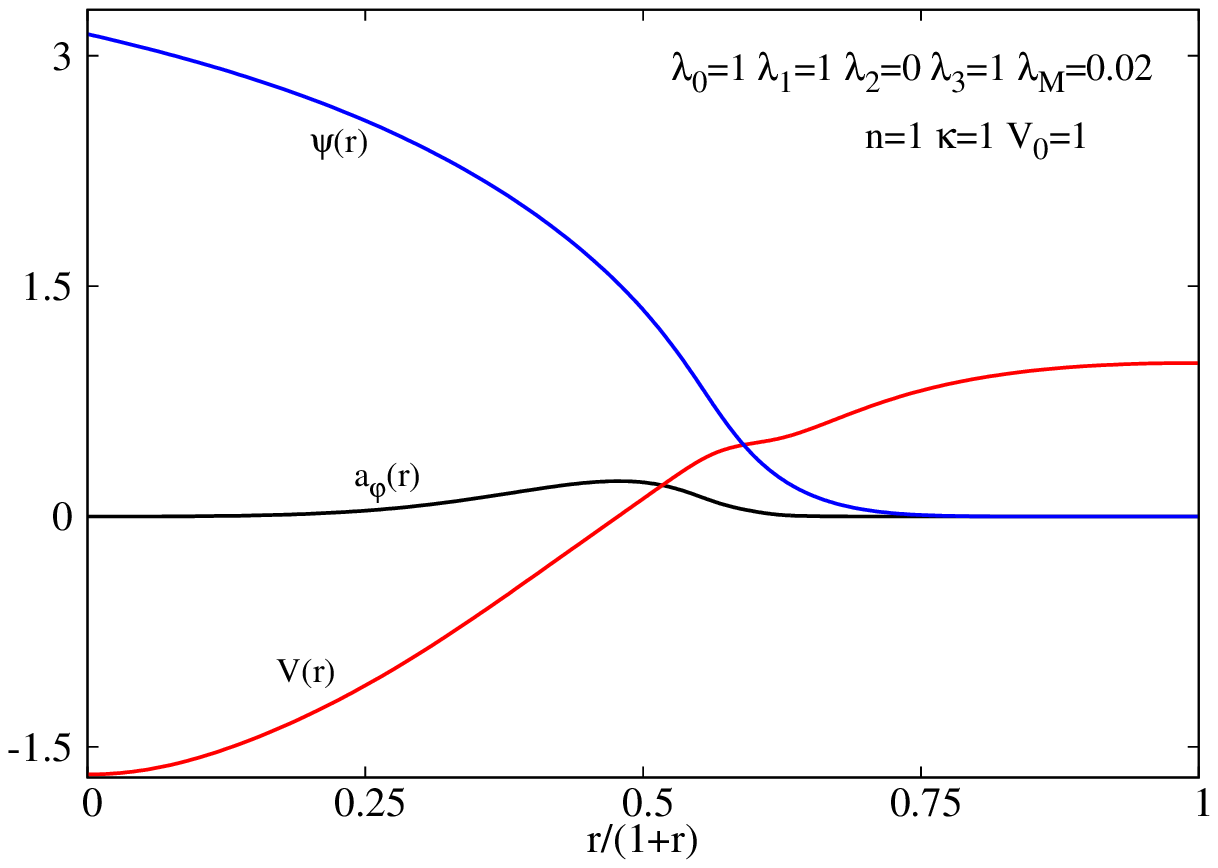}
\end{center}
\caption{
The scalar function  
$\psi$  
is shown together with the magnetic and electric gauge  potentials, $a_\varphi$ and $V$, 
for a spinning solution with $n_1=n_2=n=1$ for $\kappa=0$ (left panel) and $\kappa=1$ (right panel).
}
\label{rot1}
\end{figure}
\begin{figure}[t!]
\begin{center}
\includegraphics[height=.34\textwidth, angle =0 ]{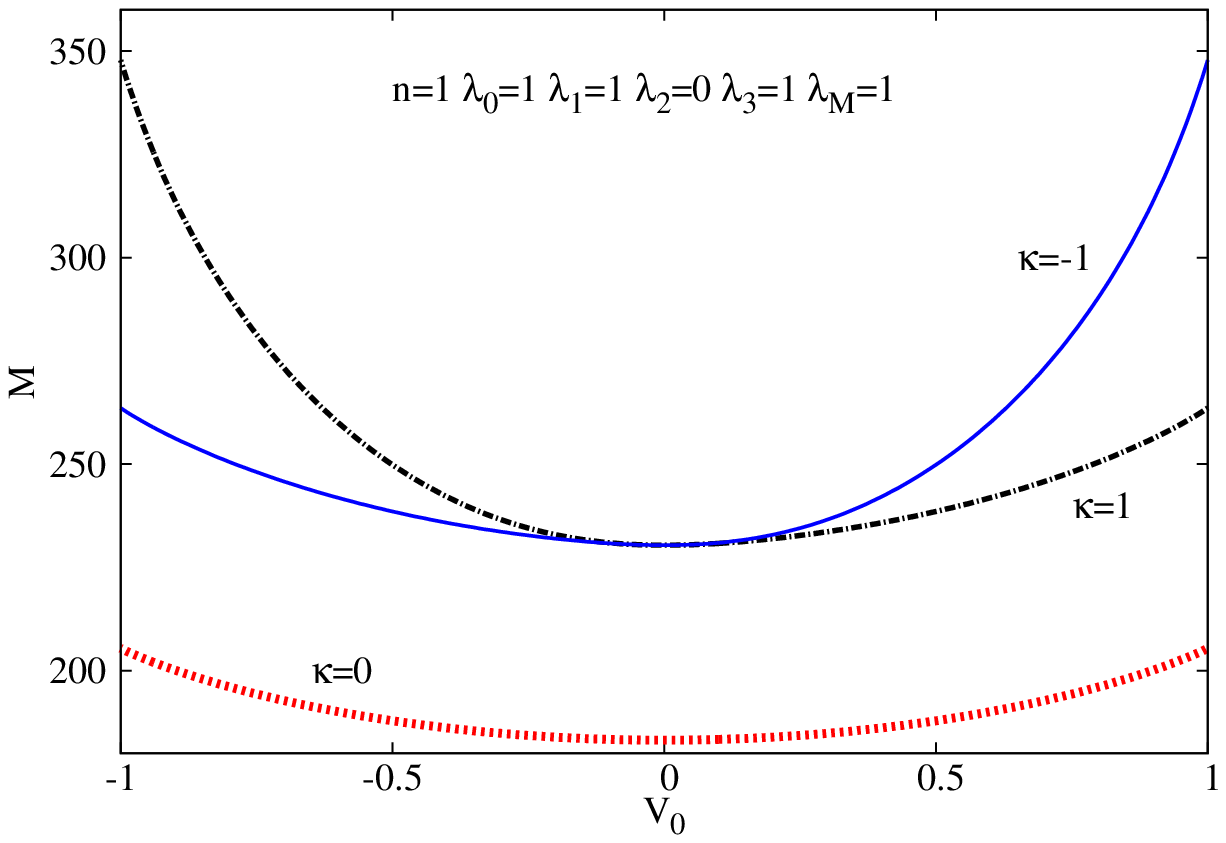}
\includegraphics[height=.34\textwidth, angle =0 ]{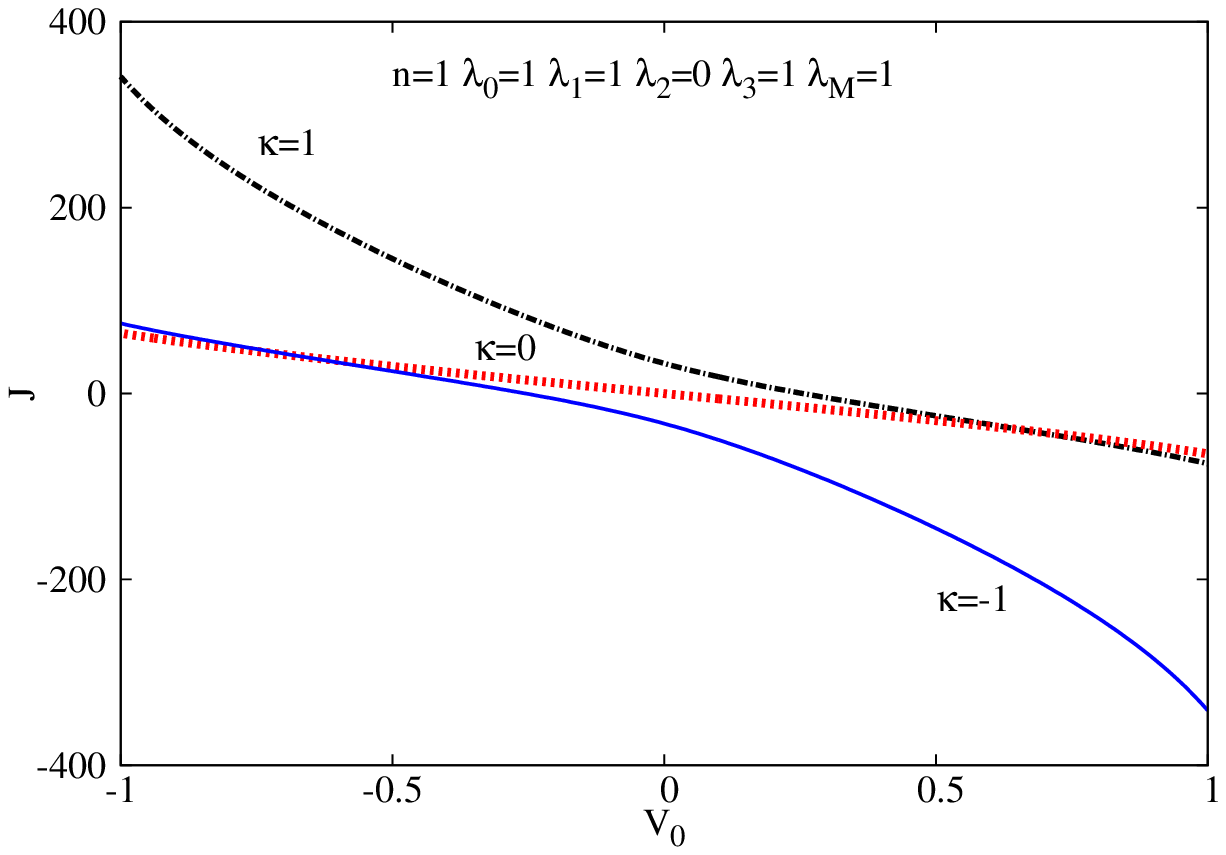}
\end{center}
\caption{
The total mass-energy $M$ 
(left panel)
and the angular momenta $J_1=J_2=J$ 
(right panel) 
are shown as functions of the 
asymptotic value of the electric potential $V_0$ for several values
of the Chern-Simons coupling constant $\kappa$ for solutions with $n_1=n_2=n=1$. 
}
\label{V0}
\end{figure}
\subsection{Electrically charged, spinning solutions}

The only rotating solutions considered in this work were found
for the enhanced symmetry Ansatz\footnote{Note that the results obtained in the static limit, Section 3.1,
strongly suggests that this limit contains already all basic features of the general case.
}
(\ref{n11})-(\ref{n12}), 
which means they have
\begin{eqnarray}
J_1=J_2=J=\frac{1}{2}\lambda_M Q_e~.
\end{eqnarray}
Also, we use the residual gauge symmetry of the model
$V\to V-\om$
to set $\om=0$ in the numerical approach.
Then,  for spinning solutions the constant $V_0$,
which fixes the asymptotic value of the electric potential,
is the only extra-input parameter as compared to the purely static case.

The profile of two typical solutions
without a CS term ($\kappa=0$, left panel) and with a CS term ($\kappa=1$, right panel)
are shown in Figure \ref{rot1}.
Note that while all other input parameters are kept constant there
(in particular the asymptotic value of the electric potential $V_0$),
the presence of a CS term leads to a rather different shape of the gauge potentials $a_\varphi(r)$ and $V(r)$.
Also,  in the $\kappa=0$ case (no CS term),
 the profiles of $a_\varphi(r)$, $\psi(r)$
are not very different as compared to the static  limit, Figure \ref{1d}.

Our numerical results indicate that
any static configuration appears to possess
rotating generalizations. As we increase
$V_0$ from zero while keeping fixed other input parameters, 
a branch of solutions forms. Along this branch, the total mass-energy $M$  increases monotonically with $V_0$.
%
The dependence of mass-energy $M$ and angular momentum $J$ on the value of the electric potential at infinity $V_0$
is shown in Figure \ref{V0}
for several values of the   
CS coupling constant $\kappa$.
As one can see,
the symmetry 
\begin{equation}
V_0 \to -V_0, ~~M \to M,~~J\to -J \ ,
\end{equation}
exists for solutions without a Chern-Simons term, $\kappa=0$, only.
Also, no upper bound appears to exist for the value of $|\kappa|$,
both the mass-energy and angular momenta increasing proportionally with $|\kappa|$.
Moreover, for $\kappa=0$, the solutions with $V_0 = 0$ have no electric field and
 correspond to static configurations discussed in Section  3.1. 
However, as expected, the angular momentum does not vanish for solutions with a CS term
which have an electric potential which vanishes asymptotically, $V_0=0$.
However, in all cases the minimal value of $M$
is approached for $V_0=0$.
Finally, let us remark on the existence of a special set of solutions with
$J=Q_e=0$,
which, for $\kappa \neq 0$ still possess a nonvanishing angular momentum/electric charge
density.

\section{Summary and outlook}
In this work, we have formulated the $SO(2)$ gauged $O(5)$ Skyrme model in $4+1$ dimensions. 
This is the gauged analogue of a previous work in Ref.~\cite{Brihaye:2017wqa}, where the gravitating
system was  studied, 
while here we consider the Maxwell dynamics instead of gravity. 
The present work is a  preliminary step towards a comprehensive investigation of the
 properties of solitons of $U(1)$ gauged Skyrmions,
in principle in all dimensions but most importantly in $3+1$.
 Here we start with $4+1$ dimensions since this is the simplest next case to $2+1$ dimensions, like which $4+1$ is an odd dimension. 
 Our ultimate aim is $ a)$ studying the charge-mass and spin-mass dependences, and $b)$ 
tracking the evolution of values of the effective baryon charge. The ``effective baryon charge'' in question is given by the lower bound on the energy of the gauged static soliton.
The solutions we seek include those supporting the global charges: electric charge and angular momentum.

We are motivated by a number of unexpected results that we have obtained in the study of an analogous model 
in $2+1$ dimensions, namely the $U(1)$ gauged planar Skyrme system,
 in which case the presence of electric charge
and spin is contingent on the presence of the Chern-Simons term in the Lagrangian.
 Specifically, we found that in $2+1$ dimensions $a)$ charge-mass and spin-mass curves are not monotonically increasing 
$as\ usual$
but rather can also decrease in some areas of the parameter space (see Section {\bf 4} of Ref.~\cite{Navarro-Lerida:2016omj}), 
and  $b)$ that in the given model solitons characterized with
continuous values of the baryon charge can exist (see Refs.~\cite{Navarro-Lerida:2018giv,Navarro-Lerida:2018siw}).

One may expect that some of the properties observed in \cite{Navarro-Lerida:2016omj,Navarro-Lerida:2018giv,Navarro-Lerida:2018siw}  
can be reproduced in the case of charged Skyrmions in $3+1$ dimensions.
But in the absence of a Chern-Simons term in $3+1$ dimensions, this does not turn out to be the case,
 as shown in our work of Refs.~ \cite{Piette:1997ny,Radu:2005jp}. While it is possible to
employ a ``new'' $U(1)$ Chern-Simons density proposed in \cite{Tchrakian:2015pka}, where it is called a Skyrme--Chern-Simons density,
this would involve the interaction with an $O(6)$ Skyrme scalar in addition to the $O(4)$ Skyrme scalar
that supports the soliton. This complication, along with the necessity to tackle a two-dimensional PDEs problem, 
is one reason we defer that $3+1$ dimensional problem and proceed in the present work to
the study of the $U(1)$ gauged $O(5)$ model in $4+1$ dimensions. 
Moreover, it is convenient that in $4+1$ dimensions the ``usual'' Chern-Simons term is available.

\medskip
The main results of this work can be summarized as follows.
First, we have established the existence of  $SO(2)$ gauged generalizations of
$O(5)$ Skyrme model introduced in Ref.~\cite{Brihaye:2017wqa}. 
(Note that here the solutions were found for a fixed Minkowski spacetime, the  gravity effects being ignored.)
Both static and spinning configurations were studied, subject to a specific Ansatz
which
reduces the problem to solving a set of PDEs.
Moreover, the Ansatz allows for an 
``enhanced symmetry'' limit that renders the residual system one-dimensional, 
depending on the radial variable only.
The static purely magnetic solutions
were studied for both the general Ansatz and the 
enhanced symmetry Ansatz. It turns out that
the numerical results show that the basic qualitative features are rather similar in both cases, as shown $e.g.$
in Figure \ref{Mq};
for example, the mass of the solutions always decreases monotonically as the gauge coupling constant increases.
In the spinning case, only enhanced symmetry configurations were studied,
 solutions with and without a Chern-Simons (CS) term being considered. 
Our numerical results  
 indicate that, different from  
the case of charged Skyrmions in $2+1$ dimensions,
the presence of a CS term
in the Lagrangian of the present model leads only to some  quantitative features,
the unusual features unveiled in
Refs.  
\cite{Navarro-Lerida:2016omj,Navarro-Lerida:2018giv,Navarro-Lerida:2018siw},
being not recovered in our study.
For example,  while in $2+1$ dimensions the 
electric charge and spin were not supported by the Skyrmion when the CS  term was absent, 
here the situation is more like 
as for gauged Skyrmions in $3+1$ dimensions
\cite{Piette:1997ny,Radu:2005jp}
or like
in the case of JZ-dyons~\cite{Julia:1975ff} where electric
charge (but not spin!) is present despite the absence of a CS term.
Also, the mass of the solutions and angular momentum always possess a monotonic dependence on the
asymptotic value of the electric potential, see  Figure \ref{V0}.

\medskip

To summarize, we conclude that
the presence of the CS term has only a quantitative effect on the $4+1$ dimensional model proposed in this paper. 
This strongly contrasts with the $2+1$ dimensional
case in \cite{Navarro-Lerida:2016omj,Navarro-Lerida:2018giv,Navarro-Lerida:2018siw},
where it had the qualitative effect of featuring
``non-standard'' mass-charge and mass-spin dependences, and, moreover, changing ``baryon number'' inside a given theory. The salient difference between the Julia-Zee (JZ) type of dyons~\cite{Julia:1975ff,Piette:1997ny,Radu:2005jp} studied here, and the Paul-Khare (PK) type dyons~\cite{Paul:1986ix} is that the former exist with or without the presence of Chern-Simons
dynamics, while the existence of the latter (PK) type is predicated on the presence of Chern-Simons dynamics in which case the Chern-Pontryagin (CP) index (in the spatial subspace) determines the electric charge (and spin) quantitatively. Given that in the model studied here the only gauge field is Abelian, the corresponding CP charge vanishes. Technically, in the case of PK dyons the presence of the CS term in the Lagrangian leads to solutions where the electric function $V(r)$ can take on a continue range of values, in contrast with the JZ type here.
 It is this property of the solutions which gives rise to the unusual mass-electric charge/spin relation and the variable ``baryon charge''
seen in~\cite{Navarro-Lerida:2016omj,Navarro-Lerida:2018giv,Navarro-Lerida:2018siw}.

To recover these properties the model at hand needs to be extended, and for
this there are two distinct possibilities available:
\begin{itemize}
\item
To extend the model to feature an $SU(2)\sim SO_{\pm}(4)$ field, 
such that the volume integral of the second Chern-Pontryagin (CP) term in the Gauss law equation (resulting from the
Maxwell equation \re{Max}), which yields the electric charge $Q_e$, does not vanish, as it does in 
the solely $U(1)$ gauged model here. This would cause the value of  $Q_e$ to depend on the CS coupling $\ka$, unlike here. 
This done, the electric charge would still get contributions even when $\ka=0$.
To change this, namely to cause $Q_e$ to be entirely dependent on $\ka$ for its support,
it would be necessary to replace $F_{\mu\nu}^2$, the kinetic term of the gauge field, with $F_{\mu\nu\rho\si}^2$. Such a model is under active consideration now.
\item
An alternative extension of the model is motivated by our study of the $SU(3)$ and $SO(5)$ gauged Higgs (YMH) model with algebra-valued Higgs
field~\cite{Navarro-Lerida:2013pua,Navarro-Lerida:2014rwa} in $3+1$ dimensions,
augmented with a Higgs--Chern-Simons (HCS) term~\cite{Tchrakian:2010ar,Radu:2011zy}. Those models, where a new Chern-Simons density (the HCS) is present, both electric charge and spin are supported.
Most importantly, they feature ``non-standard'' mass-charge and mass-spin dependenciesin~\cite{Navarro-Lerida:2016omj}. In those models, this effect has been
enabled by the larger (than $SU(2)$) gauge group.
This may signal the possibility that in the present case where the scalar matter is a Skyrme rather than a Higgs scalar, incorporating a larger target space sigma model may be useful. In this
direction, it is natural to augment the Abelian gauged $O(5)$ sigma model
with a Skyrme--Chern-Simons~\cite{Tchrakian:2015pka} density which is defined by the supplementary $O(7)$ Skyrme scalar. Compared to the above described possibility, this alternative is a
technically more challenging problem.
\end{itemize}

\section*{Acknowledgements}
The  work of E.R.  is  supported  by  the Center  for  Research  and  Development  in  Mathematics  and  Applications  (CIDMA)  
through  the Portuguese Foundation for Science and Technology (FCT - Fundacao para a Ci\^encia e a Tecnologia), 
references UIDB/04106/2020 and UIDP/04106/2020 and by national funds (OE), through FCT, I.P., 
in the scope of the framework contract foreseen in the numbers 4, 5 and 6 of the article 23,of the Decree-Law 57/2016, of August 29, changed by Law 57/2017, of July 19.  
We acknowledge support  from  the  projects  PTDC/FIS-OUT/28407/2017 
 and  CERN/FIS-PAR/0027/2019.  
 This work has further been supported by the European Union’s Horizon 2020 research and innovation (RISE) programme H2020-MSCA-RISE-2017 Grant No. FunFiCO-777740.  
The authors would like to acknowledge networking support by the COST Action CA16104

\appendix

\section{``Topological charge'' of $SO(2)$ gauged $O(5)$ model on $\R^4$}
\setcounter{equation}{0}
\renewcommand{\theequation}{A.\arabic{equation}}
What we refer to as the ``topological charge'' density of a gauged Skyrme (sigma model) system is the density that results from the deformation of the topological charge density of the sigma model, {\it prior to gauging}. While the topological charge of the latter presents a
lower bound on the energy of the ungauged sigma model, the latter presents the energy lower bound for the gauged system. As such, it is not strictly speaking a topological charge, as is the case for gauged Higgs systems.

The lower bound on the energy of a given sigma model in the appropriate dimensions is given by a topological~\cite{Manton:2004tk} charge density. Such charge densities are not explicitly {\it total divergence}, in contrast to the case of Higgs 
models~\cite{Tchrakian:2010ar}, but when the sigma model scalar is expressed in a parametrisation that is compliant with the sigma model constraint, they become explicitly {\it total divergence}. We refer to the charge densities in the generic parametrisation
as {\it essentially total divergence}. In the context of the present work, the sigma models in question are the $O(D+1)$ Skyrme models on $\R^D$.

When it comes to gauging Skyrme models, the situation strictly differs from the (gauged~\footnote{The gauge decoupled Higgs models, referred to as Goldstone models in \cite{Tchrakian:2010ar}, do support topologically stable solitons.}) Higgs models~\cite{Tchrakian:2010ar},
where the charge densities supplying the lower bounds on the energy densities are all descended from Chern-Pontryagin densities in some higher dimension, and are topological densities. The gauge group of a Higgs model is fixed by the representation of the Higgs scalar, in
which the topological charge is encoded~\cite{Arafune:1974uy,Tchrakian:2002ti}. Gauging a Skyrme model with the requirement that a charge density giving a lower bound on the energy density be defined, contrasts starkly with the definition of the corresponding density of a
Higgs model. The physical requirements that the charge density of a gauged Skyrmion must satisfy, $a)$ that it be {\it (essentially) total divergence} to enable the evaluation of the charge integral as a surface integral in terms of the asymptotic fields, and,
$b)$ that it be gauge invariant. A prescription for achieving such a definition was given in \cite{Tchrakian:1997sj}.

The definition of the topological charge density of an $SO(N)$ gauged Skyrme scalar $\f^a=(\f^{a},\f^{D+1})\ ,\ \ a=1,2,\dots,D$ on $\R^D$ given in ~\cite{Tchrakian:1997sj} and \cite{Tchrakian:2015pka} relies on the relation between the winding number density $\vr_0$, prior
to gauging, which is {\it effectively total divergence} but is {\it gauge variant}, and the density $\vr_G$ defined by replacing all the partial derivatives in $\vr_0$ by covariant derivatives. While $\vr_0$ is {\it effectively total divergence} and is {\it gauge\ variant},
$\vr_G$ is {\it gauge invariant} but is not {\it total divergence}. The physical charge density must be both {\it gauge invariant} and {\it total divergence}. The definition of $\vr_G$ follows from that of $\vr_0$, by formally replacing all partial derivatives by
the $SO(N)$, covariant derivatives.

The generic expression
\bea
\vr&=&\vr_0+\pa_i\Om_i[{\cal A},\f]\label{top14}\\
&=&\vr_G+W[{\cal F},{\cal D}\f] \ ,  \label{top24}
\eea
defines the topological charge density where $\Om_i[{\cal A},\f]$, like $\vr_0$, is {\it gauge variant} while $W[{\cal F},{\cal D}\f]$, like $\vr_G$ is $gauge\ invariant$. \re{top14} and \re{top24} are equivalent definitions for $\vr$, which as required is
both {\it gauge invariant} and {\it total divergence}.

Clearly, the definition \re{top14}-\re{top24} depends on the gauging prescription employed and here, in contrast with Higgs models, the gauge group can be chosen to be $SO(N)$, for all $N$ in the range $D\ge N\ge 2$. Thus, at most $D$ components of the $D+1$
component Skyrme scalar are gauged with $SO(D)$, down to only two of the components with (Abelian) $SO(2)$. For example in \cite{Piette:1997ny,Radu:2005jp}, the $O(4)$ Skyrme system on $\R^3$ is gauged with $SO(2)$, while in \cite{Arthur:1996np} this system is gauged with
$SO(3)$. Numerous such examples are presented in \cite{Tchrakian:2015pka}, for models on $\R^D$, $D=2,3,4,5$.

In the present work, we are concerned with the $O(5)$ model on $\R^4$, with the Abelian gauging prescription \re{coval}, \re{covA} and \re{cov3}, $i.e.$, gauging two pairs of Skyrme scalars, in contrast to the example given in \cite{Tchrakian:2015pka} where only one
pair is gauged. Our choice for gauging two pairs here is driven by our desire of having a gauging prescription that is symmetric in $a=1,2$ and $a=3,4$ that enables the enhanced radial symmetry of the bi-azimuthal symmetry. For this reason it is convenient to start with
the maximal $SO(4)$ gauging and proceeding to the desired sub-gauging prescriptions by group contraction. We describe these two steps in the next two subsections.

\subsection{Topological charge of $SO(4)$ gauged $O(5)$ Skyrme system}
We denote the $O(5)$ Skyrme scalar $\f^a$, $a=a',5\,\ \ a'=1,2,3,4$ and the densities $\vr_0$ and $\vr_G$ appearing in \re{top14}-\re{top24} are
\bea
\varrho_0 &=&\vep_{ijkl} \vep^{abcde}\pa_i\f^a\pa_j\f^b\pa_k\f^c\pa_l\f^d\f^e\ ,\label{vr0}\\
\varrho_G &=&\vep_{ijkl} \vep^{abcde}{\cal D}_i\f^a{\cal D}_j\f^b{\cal D}_k\f^c{\cal D}_l\f^d\f^e \ , \label{vG0}
\eea
where calligraphic ${\cal D}$ and ${\cal A}$ are used to denote the $SO(4)$ gauged covariant derivative and connections. This is to distinguish these quantities from $D$ and $A$ of  the $SO(2)$ gauging in the main body of the work. Thus the $SO(4)$
connection and curvature
are expressed by $({\cal A}_i^{a'b'},{\cal F}_{ij}^{a'b'})$, and the gauging prescription by the covariant derivative
\[
{\cal D}_i\f^{a'}=\pa_i\f^{a'}+{\cal A}_i\f^{a'}\ ,\ {\cal D}_i\f^5=\pa_i\f^5,
\]
where ${\cal A}_i\f^{a'}={\cal A}_i^{a'b'}\f^{b'}$ and $i=1,2,3,4$.

The quantities  $\Om_i[{\cal A},\f],W[{\cal F},{\cal D}\f]$ in \re{top14}-\re{top24} are given in \cite{Tchrakian:1997sj}-\cite{Tchrakian:2015pka} to be
\bea
\Om_i&=&3!\,\vep_{ijkl}\vep^{a'b'c'd'}\f^5\bigg\{\pa_j\left[{\cal A}_l^{a'b'}\f^{c'}(\pa_k\f^{d'}+\frac12{\cal A}_k\f^{d'})\right]+\nonumber\\
&&\qquad\qquad\qquad\ \ \ \ +\frac14\left(1-\frac13(\f^5)^2\right){\cal A}_l^{a'b'}\left[\pa_j{\cal A}_k^{c'd'}+\frac23({\cal A}_j{\cal A}_k)^{c'd'}\right]\bigg\} \ ,  \label{Om4}\\
W&=&3!\,\vep_{ijkl}\vep^{a'b'c'd'}\left\{\frac{1}{16}\f^5\left(1-\frac13(\f^5)^2\right){\cal F}_{ij}^{a'b'}{\cal F}_{kl}^{c'd'}+\frac12\,{\cal F}_{ij}^{a'b'}\f^{c'}{\cal D}_k\f^{d'}\pa_l\f^5\right\} \ ,  \label{W4}
\eea
in which $\Om_i[{\cal A},\f]$ is manifestly gauge variant, displaying the (Euler)--Chern-Simons density which is typical in all even dimensions, and $W[{\cal F},{\cal D}\f]$ which is manifestly gauge invariant.
Clearly, the Euler density can be added to the definitions of both \re{Om4} and \re{W4}, but this is unnecessary here, especially since we do not anticipate the introduction of a potential term in the Lagrangian.

What is important to realise here is that the gauge variant density \re{Om4}, consists $exclusively$ of gauge variant elements. This feature, which occurred in the $2$ and $3$ dimensional cases seen in
\cite{Tchrakian:1997sj}-\cite{Tchrakian:2015pka}, is not a general feature 
in all dimensions. In dimensions $D\ge 4$, there is the freedom to remove the total divergence part of the second term in \re{W4} and placing it in \re{Om4}. The result is again a gauge invariant definition of the topological charge.
$D\ge 4$, the definitions can be altered such that the gauge variant density consists of $both$ gauge variant and gauge invariant terms. In this redefinition, $\Om_i$ remains gauge variant, while $W$ remains gauge invariant as required. 

By removing a total divergence term in \re{W4} and placing it in \re{Om4}, we have the more aesthetic definitions for the densities
\bea
\hat \Om_i&=&3!\,\vep_{ijkl}\vep^{a'b'c'd'}\f^5\bigg\{\frac12{\cal F}_{kl}^{c'd'}\f^{a'}{\cal D}_j\f^{b'}+\pa_j\left[{\cal A}_l^{a'b'}\f^{c'}(\pa_k\f^{d'}+\frac12{\cal A}_k\f^{d'})\right]+\nonumber\\
&&\qquad\qquad\qquad\ \ \ \ +\frac14\left(1-\frac13(\f^5)^2\right){\cal A}_l^{a'b'}\left[\pa_j{\cal A}_k^{c'd'}+\frac23({\cal A}_j{\cal A}_k)^{c'd'}\right]\bigg\} \ ,  \label{Om4f}\\
\hat W&=&3!\,\vep_{ijkl}\vep^{a'b'c'd'}\f^5\left\{\frac{1}{24}(\f^5)^2{\cal F}_{ij}^{a'b'}{\cal F}_{kl}^{c'd'}+\frac12\,{\cal F}_{ij}^{a'b'}{\cal D}_{[k}\f^{c'}{\cal D}_{l]}\f^{d'}\right\} \ . \label{W4f}
\eea
Apart from its aesthetic appearance, the density $\hat W$, \re{W4f}, is necessary for the statement of the relevant Belavin inequalities in Appendix {\bf B} below.

\subsection{Group contraction}
In \re{Om4}-\re{W4} and \re{Om4f}-\re{W4f}, ${\cal A}_i^{a'b'}$ is the $SO(4)$ connection, ${\cal F}_{ij}^{a'b'}$ is the curvature, and  ${\cal D}_i\f^{a'}=\pa_i\f^{a'}+{\cal A}_i\f^{a'}$ is the covariant derivative,
with $a'=1,2,3,4$. In the notation of \re{coval}-\re{covA}, $a'=\al,A;\ \ \al=1,2;\ A=3,4$.

We now contract the gauge group $SO(4)$ by setting the components of the connection ${\cal A}_i^{a'b'}=({\cal A}_i^{\al\bt},{\cal A}_i^{AB},{\cal A}_i^{\al A})$ to 
${\cal A}_i^{\al\bt}=A_i\vep^{\al\bt}\ ,\ \ {\cal A}_i^{AB}=B_i\vep^{AB}$, and ${\cal A}_i^{\al A}=0$, where $A_i^{\al\bt}$ and $B_i^{AB}$ are now two $SO(2)$ connections inside $SO(4)$. The corresponding curvatures are
$F_{ij}=\pa_{[i}A_{j]}$, $G_{ij}=\pa_{[i}B_{j]}$ and $F_{ij}^{\al A}=0$.
The covariant derivative ${\cal D}_i\f^{a'}=({\cal D}_i\f^\al,{\cal D}_i\f^A)$ now splits up into
\bea
{\cal D}_i\f^{\al}&=&\pa_i\f^{\al}+A_i(\vep\f)^{\al}  \ ,  \label{cocconal}\\
{\cal D}_i\f^{A}&=&\pa_i\f^{A}+B_i(\vep\f)^{A}  \ . \label{cocconA}
\eea
Finally, identifying~\footnote{Alternatively, setting $B_i=0$, \re{Om4f}-\re{W4f} leads to the corresponding topological charge density displayed in \cite{Tchrakian:2015pka}, where only one pair of Skyrme scalars is gauged with $SO(2)$.} the two Abelian fields
$A_i=B_i$, \re{cocconal}-\re{cocconA} we have the desired gauging prescription \re{coval}-\re{covA}.

As a result of this group contraction, the topological charge densities following from \re{Om4f}-\re{W4f} are expressed by
\bea
\hat\Om_i&=&3!\,\vep_{ijkl}\,\f^5A_l\left\{-\frac13(\f^5)^2F_{jk}+2\left(\vep^{AB}\pa_j\f^{A}\pa_k\f^{B}+ \vep^{\al\bt}\pa_j\f^{\al}\pa_k\f^{\bt}\right)\right\}  \ , \label{Om2}\\
\hat W&=&3!\,\vep_{ijkl}\,\f^5\,F_{ij}\left\{\left(1-\frac13(\f^5)^2\right)F_{kl}+2\left(\vep^{\al\bt}D_k\f^{\al}D_l\f^{\bt}+\vep^{AB}D_k\f^{A}D_l\f^{B}\right)\right\}  \ ,  \label{W2}
\eea
where $D_i\f^{\al}$ and $D_i\f^{A}$ are now given by \re{coval} and \re{covA} respectively, and $F_{ij}=\pa_iA_j-\pa_jA_i$.

Inserting $\hat\Om_i$ and $\hat W$ given by \re{Om2}-\re{W2} into \re{top14}-\re{top24} defines the topological charge density for the $SO(2)$ gauged $O(5)$ model studied here.

\subsection{Charge integrals}
We adopt the definition of the topological charge density \re{top14} with $\Om_i$ being given by \re{Om2}, and we denote the second term by $\vr_1=\pa_i\hat\Om_i$. This term is manifestly total divergence, while the first term in \re{top14}, namely $\vr_0$ defined by
\re{vr0} is not manifestly total divergence but becomes such when a constraint compliant parametrisation of the scalar $\f^a$ satisfying \re{constr} is employed. For this purpose, we adopt the parametrisation
\be
\Psi_1=\sin f\,\sin g\ ,\quad\Psi_2=\sin f\,\cos g\ ,\quad\Psi_3=\cos f\,.\label{constr1}
\ee
In terms of the functions $f(\rho,\si)$ and $g(\rho,\si)$ (with $\rho=r\sin\ta$, $\si=r\cos\ta$), $\vr_0$ reduces to the antisymmetric product
\be
\vr_0=2\cdot 3!\,\frac{n_1n_2}{\rho\si}\,\pa_{[\rho}F\,\pa_{\si]}G\,,\label{vr01}
\ee
where $F(\rho,\si)$ and $G(\rho,\si)$ are the functions
\be
F=\cos f+\frac23\cos^3f-\frac35\cos^5f\ ,\quad G=\sin^2 g\,.\label{FG}
\ee
Denoting $(\rho,\si)=y_i \ , i=1,2$, the volume integral of $\vr_0$ can be cast in the form
\bea
\int\,\vr_0\,d^4x&=&(2\pi)^2n_1n_2\int\vep_{ij}\pa_iF\pa_jG\,d^2y\nonumber\\
&=&\frac12(2\pi)^2n_1n_2\int\vep_{ij}(F\stackrel{\leftrightarrow}\pa_j G)\,ds_i\,.\label{sirho0}
\eea
It is interesting to point out here that in evaluating the Stokes integral \re{sirho0}, instead of taking $0\le\ta\le\frac{\pi}{2}$ one can take the limits $0\le\ta\le\frac{m\pi}{2}$, with $m$ an integer. For even $m$, the solutions should be Skyrme-antiSkyrme
as in Yang-Mills.

The corresponding integral of the term $\vr_1$ can also be evaluated using Stokes theorem, since in that case this density is manifestly total divergence in terms of the functions $(f, g)$
\bea
\int\,\vr_1\,d^4x&=&(2\pi)^2\int\vep_{ij}\bigg\{\frac23\Psi_3^3\left[(a_{(1)}-n_1)\pa_ja_2-(a_{(2)}-n_2)\pa_ja_1\right]\nonumber\\
&&\qquad\qquad\qquad+2\Psi_3\left[n_1(a_{(2)}-n_2)\pa_j\Psi_1^2-n_2(a_{(1)}-n_1)\pa_j\Psi_2^2\right]\bigg\}\,ds_i\,,\label{sirho1}
\eea
where the volume integral is evaluated by applying Stokes' Theorem.

\section{The Belavin inequalities and the models}
\setcounter{equation}{0}
\renewcommand{\theequation}{B.\arabic{equation}}
We establish the Belavin inequalities for the $SO(4)$ gauged system, from which follow the corresponding inequalities pertaining to the gauge-contracted systems, in particular those giving the lower bound on the
static Hamiltonian of the Lagrangian \re{action} of the $SO(2)$ gauged model studied here.

The Belavin inequalities are most conveniently derived from definition \re{top24} of the topological charge, given by $\hat W$

Consider now the inequalities
\bea
\left|\f^5{\cal F}_{ij}^{a'b'}-\frac{1}{2!^2}\vep_{ijkl}\vep^{a'b'c'd'}{\cal F}_{kl}^{c'd'}\right|^2&\ge&0 \ ,
\label{1}\\
\left|\f^5{\cal F}_{ij}^{a'b'}-\frac{1}{2!^2}\vep_{ijkl}\vep^{a'b'c'd'}{\cal D}_{[k}\f^{c'}{\cal D}_{l]}\f^{d'}\right|^2&\ge&0  \ , \label{2}\\
\left|{\cal D}_{[i}\f^{a}{\cal D}_{j]}\f^{b}-\frac{1}{2!^2}\vep_{ijkl}\vep^{abcde}{\cal D}_{[k}\f^{c}{\cal D}_{l]}\f^{d}\f^e\right|^2&\ge&0\ ,
\quad a=a',5
\label{3}
\eea
The inequalities \re{1}-\re{3} yield
\bea
(\f^5)^2(1+(\f^5)^2)\left|{\cal F}_{ij}^{a'b'}\right|^2&\ge&\frac14\vep_{ijkl}\vep^{a'b'c'd'}(\f^5)^3{\cal F}_{ij}^{a'b'}{\cal F}_{kl}^{c'd'}  \ , \label{1y}\\
(\f^5)^2\left|{\cal F}_{ij}^{a'b'}\right|^2+\left|{\cal D}_{[i}\f^{a'}{\cal D}_{j]}\f^{b'}\right|^2&\ge&
\frac12\vep_{ijkl}\vep^{a'b'c'd'}\f^5{\cal F}_{ij}^{a'b'}{\cal D}_{[k}\f^{c'}{\cal D}_{l]}\f^{d'}\ , \label{2y}\\
\left|{\cal D}_{[i}\f^{a}{\cal D}_{j]}\f^{b}\right|^2&\ge&\vr_G \ . 
\label{3y}
\eea

Adding $\frac16$ times \re{1y} to \re{2y} and \re{3y}, the right hand sides yield the ``topological charge'' density
\[
\vr=\vr_G+\hat W[{\cal F},{\cal D}\f]\,,
\]
defined  by \re{top24}, with $W$ there replaced by $\hat W$ in \re{W4f}.


Concerning the left hand side of that inequality, this can be cast into the form $$c_1\left|{\cal F}_{ij}^{a'b'}\right|^2+c_2\left|\vf_{ij}^{a'b'}\right|^2  \ , $$ ($c_1,\ c_2>0$) by simply adding positive definite quantities, recognising also that in \re{W4f}, the quantity
$\frac12\f^5\left(1-\frac13(\f^5)^2\right)$ is always positive.

Thus, after the group contraction described in Section {\bf A.1} with ${\cal F}_{ij}^{a'b'}\to F_{ij}^{a'b'}$ and $\vf_i^a\to\f_i^a$, the static energy density functional pertaining to \re{action} is the bounded from below by $\vr_G[\f^a]$  plus $\hat W$ given by \re{W2}.
Clearly, the positive definite quadratic and sextic kinetic Skyrme terms in \re{action} can be added without invalidating the bound.

\begin{small}

\end{small}

\end{document}